\documentclass[australian,english,prl, singlespace, twocolumn]{revtex4-1}
\usepackage[T1]{fontenc}
\usepackage[latin9]{inputenc}
\usepackage{color}
\usepackage{amsmath}
\usepackage{amssymb}
\usepackage{graphicx}

\makeatletter

\newcommand*\LyXThinSpace{\,\hspace{0pt}}

\@ifundefined{definecolor}
 {\usepackage{color}}{}
\makeatother

\makeatother

\usepackage{babel}
\begin{document}
\title{Bipartite Leggett-Garg and macroscopic Bell inequality violations
using cat states: distinguishing weak and deterministic macroscopic
realism }
\author{Manushan Thenabadu and M. D. Reid$^{1}$}
\affiliation{$^{1}$ Centre for Quantum Science and Technology Theory, Swinburne
University of Technology, Melbourne 3122, Australia}
\begin{abstract}
We consider tests of Leggett-Garg's macrorealism and of macroscopic
local realism, where for spacelike separated measurements the assumption
of macroscopic noninvasive measurability is justified by that of
macroscopic locality. We give a mapping between the Bell and Leggett-Garg
experiments for microscopic qubits based on spin $1/2$ eigenstates
and gedanken experiments for macroscopic qubits based on two macroscopically
distinct coherent states (cat states). In this mapping, the unitary
rotation of the Stern-Gerlach analyzer is realized by an interaction
$H=\Omega\hat{n}^{4}$ where $\hat{n}$ is the number of quanta. By
adjusting the time of interaction, one alters the measurement setting.
We thus predict violations of Leggett-Garg and Bell inequalities in
a macroscopic regime where  coarse-grained measurements $\hat{M}$
need only discriminate between two macroscopically distinct coherent
states. To interpret the violations, we distinguish
between different definitions of macroscopic realism. Deterministic
macroscopic local realism (dMR) assumes a definite outcome for the
measurement $\hat{M}$ prior to the unitary rotation created by the
analyser, and is negated by the violations. Weak macroscopic realism
(wMR) assumes a definite outcome for systems prepared in a superposition
$\psi_{pointer}$ of two macroscopically-distinct eigenstates of $\hat{M}$,
after the unitary rotation. We find that wMR can be viewed as consistent
with the violations. A model is presented, in which wMR holds, and
for which the macroscopic violations emerge over the course of the
unitary dynamics. Finally, we point out an EPR-type paradox,
that a weak macro-realistic description for the system prior to the
measurement $\hat{M}$ is inconsistent with the completeness of quantum
mechanics.
\end{abstract}
\maketitle

\section{Introduction}

Bell's theorem constrains the predictions of all local hidden variable
theories to values governed by inequalities \cite{bell-brunner-rmp,CShim-review,Bell-2,det-bell}.
These inequalities can be violated by quantum mechanics, giving a
falsification of both local realism and local causality. Bell tests
involve measurements made on two separated particles, performed on
timescales such that the two measurement events are spacelike separated.
At each location, there is a choice between two measurement settings.
To date, almost all Bell tests have examined microscopic systems.
Bell violations have been predicted for large numbers of particles
\cite{meso-bell-higher-spin-cat-states-bell,cv-bell,cat-bell,mdr-mlr,macro-bell-additional,cv-bell-macro-ali},
but, in almost all cases, the measurements for one of the settings
require a resolution of a few particles, or at the level of Planck's
constant.

By contrast, Leggett and Garg derived inequalities that do not require
such fine resolution of measurement \cite{legggarg-1}. Leggett-Garg
inequalities are designed to falsify the assumption of \emph{macrorealism}
(M-R) for systems that at certain times are in a superposition (e.g.
$\psi_{\alpha}\sim|a\rangle+|d\rangle$) of two  macroscopically
distinct states \cite{s-cat-1}. M-R combines two premises: Macroscopic
realism (MR) asserts that the system must actually be in one or other
of those two states at the given time. The second premise, macroscopic
noninvasive measurability (NIM), asserts the existence of a measurement
$\hat{M}$ that can distinguish between the two macroscopic states,
with negligible effect on the subsequent dynamics (at a macroscopic
level). Macroscopic realism implies a predetermined value $\lambda_{M}$
for the outcome of the measurement $\hat{M}$ \cite{legggarg-1}.
Since the two states $|a\rangle$ and $|d\rangle$ can be distinguished
by a measurement $\hat{M}$ allowing a macroscopic uncertainty, the
hidden variable $\lambda_{M}$ need not specify the results of measurements
to a resolution of $\hbar$. This was recognised by Leggett and Garg
who assigned the values $\lambda_{M}=\pm1$ for the two outcomes of
$\hat{M}$, in order to derive inequalities \cite{legggarg-1}.

Leggett-Garg inequalities have been reported violated for a wide range
of systems \cite{emary-review,jordan_kickedqndlg2-1,weak-solid-state-qubits,NSTmunro-1,massiveosci-1-1,manushan-cat-lg,weak-hybrid,leggett-garg-recent,macro-bell-lg,nst}.
A complication for the interpretation of the violations has been the
justification of the NIM premise for practical measurements \cite{leggett-garg-recent}.
Different strategies are used, including (in a recent macroscopic
superconducting experiment) a control experiment which quantifies
the amount of disturbance induced by the measurement, if the system
is indeed in one of the states $|a\rangle$ or $|d\rangle$ \cite{NSTmunro-1}.
An alternative strategy (that does not assume the system to be in
one of the fully specified quantum states) is to violate bipartite
Leggett-Garg or macroscopic Bell inequalities \cite{macro-bell-lg,weak-hybrid,cv-bell-macro-ali}.
Here, the NIM premise is replaced by that of macroscopic locality.
Recent work gives predictions for such violations using NOON and multi-component
cat states \cite{macro-bell-lg}. The question becomes how to interpret
such violations. In particular, we ask whether the assumption of the
macroscopic hidden variable $\lambda_{M}$ for the coarse-grained
measurement $\hat{M}$ can be negated.

In this paper, we give full details of the results presented in the
Letter \cite{letter}, which analyses this question. To facilitate
the analysis, we give predictions of violations of macroscopic Bell
inequalities and of bipartite Leggett-Garg inequalities  for the
conceptually simple case where a system at certain times is found
to be in a superposition of two macroscopically distinct coherent
states, $|\alpha\rangle$ and $|-\alpha\rangle$ (cat states), as
$\alpha\rightarrow\infty$. Similar to the results of Ref. \cite{macro-bell-lg},
the violations are obtained using only measurements $\hat{M}$ which
discriminate between the two coherent states, thereby allowing macroscopic
uncertainties.

In order to interpret the violations, we are first careful to define
macroscopic realism in the weakest (i.e. most minimal) sense, as \emph{weak
macroscopic realism} (wMR), the premise that a macroscopic hidden
variable $\lambda_{M}$ specifies the outcome of the\emph{ }coarse-grained
pointer measurement $\hat{M}$. In contrast to many previous macrorealism
tests, we find it is not necessary to include in the definition of
macroscopic realism that the system be in one or other of two specific
quantum states (e.g. $|\alpha\rangle$ and $|-\alpha\rangle$), or
even that the system be in one or other of two unspecified quantum
states. The violations given in this paper therefore provide strong
tests of both macrorealism (M-R) and macroscopic local realism (MLR),
and also of macroscopic local causality (MLC).

In this paper, we further weaken the definition of weak macroscopic
realism, by specifying $\hat{M}$ to be a pointer measurement, and
the distinct states $|a\rangle$ and $|d\rangle$ to be states with
a definite outcome for the (coarse-grained) $\hat{M}$. In other words,
the assumption of wMR asserts the validity of the macroscopic hidden
variable $\lambda_{M}$ for the state $\psi_{\alpha}$ prepared \emph{after}
a unitary transformation $U$ that determines the choice of measurement
setting. We show that the violations of M-R and MLR/ MLC given in
this paper can be viewed \emph{consistently} with wMR.

On the other hand, we show that the violations falsify \emph{deterministic
macroscopic realism} (dMR), which specifies well-defined values $\lambda_{M}$
\emph{prior} to such a unitary transformation. The assumption of dMR
also includes the assumption of macroscopic locality (ML), which states
that the value of $\lambda_{M}$ cannot be changed by spacelike separated
measurement events at another site. The premise of dMR implies the
validity of \emph{two} hidden variables, $\lambda_{M_{\theta}}$ and
$\lambda_{M_{\theta'}}$, ascribed to the system \emph{simultaneously}
(prior to $U$) to predetermine the outcome of two pointer measurements,
$\hat{M}_{\theta}$ and $\hat{M}_{\theta'}$, and is negated by the
violation of the macroscopic Bell inequalities.

We find that the consistency of the violations with wMR is possible
because of the unitary dynamics associated with the choice of measurement
setting. The unitary rotation $U=e^{-iHt/\hbar}$ that determines
the measurement setting transforms the system from one pointer superposition
to another, over a time interval $\Delta t$. The system is therefore
not considered to be in a superposition of eigenstates of both pointer
measurements, $\hat{M}_{\theta}$ and $\hat{M}_{\theta'}$, simultaneously.
In our analysis, the transformation $U$ that determines the measurement
setting is given by a nonlinear Hamiltonian $H=\Omega\hat{n}^{4}$
where $\hat{n}$ is the mode number operator. The dynamics leading
to the macroscopic violations can therefore be observed over the interval
$\Delta t$. Recent work predicts violations of Bell inequalities
at a microscopic level for the dynamical trajectories of two entangled
particles \cite{bell-trajectories}. This is consistent with our conclusion
that deterministic macroscopic realism fails, since the choice of
measurement setting involves dynamics, as given by $U$.

There are open questions. While we show consistency with wMR, the
results of this paper neither falsify nor validate this premise. 
The validation of this premise might be expected,  in view of the
known emergence of classicality with coarse-grained measurements \cite{coarse-3,coarse-peres,kofler-bruck-leggett-garg-coarse}.
The concept might also be consistent with recent proposals to interpret
quantum mechanics using theories based on multiple interacting classical
worlds \cite{mhall}, or with an epistemic restriction of order Planck's
constant \cite{budiyono,bartlett-gaussian,macro-pointer,q-trajectories-drummond},
and may also be consistent with recent interpretations of the double-slit
experiment \cite{double-slit-Aharonov}.

However, it is clear from the violations of the Leggett-Garg inequalities
presented in this paper (and elsewhere) that if wMR holds, then the
dynamics associated with the macroscopic hidden variables $\lambda_{M}$
\emph{cannot} be given by classical mechanics. \emph{If} wMR holds,
we show for the tests of this paper that the violations of the Leggett-Garg
inequalities would therefore arise from the breakdown of the NIM assumption.
We explain how this breakdown arises (in the bipartite model) from
quantum nonlocality, the new feature being that there is a \emph{macroscopic}
nonlocality (associated with macroscopically distinct\emph{ }qubits
$|\alpha\rangle$ and $|-\alpha\rangle$), which we show emerges over
the timescales associated with the unitary interactions $U$ at \emph{both}
sites. While the distinction between predictions for the superposition
$\psi_{\alpha}$ and the classical mixture $\rho_{mix}$ of the two
macroscopically distinct states is negligible (of order $\hbar e^{-|\alpha|^{2}}$),
a \emph{macroscopic} difference emerges  over the course of the
dynamics corresponding to the unitary rotations.

From an alternative point of view, we summarise in this paper that
if one does postulate the validity of wMR, then inconsistencies arise.
This can be shown in the form of an EPR-type paradox \cite{epr-1},
similar to that given in earlier papers \cite{macro-coherence-paradox,eric_marg-1,irrealism-fringes,macro-pointer}.
If one assumes wMR, then the assumption that the system is in one
of two \emph{quantum} states, such as $|\alpha\rangle$ or $|-\alpha\rangle$,
can be negated. Hence, wMR is inconsistent with the completeness of
quantum mechanics, which gives predictions at the level of $\hbar$.

\textbf{\emph{Layout of paper}}: The paper is organised as follows.
In Section II, we review and extend previous work to give derivations
of bipartite Leggett-Garg and macroscopic Bell inequalities.  \emph{}The
macroscopic Bell inequalities are derived assuming \emph{deterministic
macroscopic (local) realism} or \emph{macroscopic local causality},
both of which are negated by violations of the inequalities given
in this paper.\emph{}

In Section III, we demonstrate a mapping between the spin-$1/2$
Bell experiments performed on microscopic qubits $|\uparrow\rangle$
and $|\downarrow\rangle$, and a gedanken experiment performed on
macroscopic qubits, $|\alpha\rangle$ and $|-\alpha\rangle$. The
unitary rotation of the polariser or spin analyser is mapped to a
transformation $|\alpha\rangle\rightarrow\cos\theta|\alpha\rangle+i\sin\theta|-\alpha\rangle$,
which is realized for $\theta=0,\pi/8,\pi/4$ and $3\pi/8$ by an
evolution $U=e^{-iHt/\hbar}$ at certain times $t$ \cite{manushan-cat-lg}.
We thus show that macroscopic Bell inequalities are violated for spacelike
separated systems, $A$ and $B$, prepared in an entangled two-mode
cat state $|\psi\rangle\sim|\alpha\rangle|-\beta\rangle-|-\alpha\rangle|\beta\rangle$.
The mapping relies on the orthogonality of the coherent states,
which strictly holds only for $\alpha$, $\beta\rightarrow\infty$.
In our explicit model, we give full calculations, and compute the
values of $\alpha$ and $\beta$ for which experiments could be performed.

In Section IV, we predict violations of Leggett-Garg's macrorealism,
for both single and bipartite cat-state systems, following along the
lines of Ref. \cite{manushan-cat-lg}. In the Bell test, the angle
$\theta$ is the measurement setting. In the Leggett-Garg test, the
angle $\theta$ gives the time $t$ of evolution between measurements
$\hat{M}$ made at different times. The dynamics associated with
the unitary evolution $U=e^{-iHt/\hbar}$ can be visualised by plotting
the $Q$ function \cite{Husimi-Q-1}. We explain \emph{weak macroscopic
realism} (wMR) in Sections III.E and also in Section VI. Evaluation
of the $Q$ function dynamics is given in Sections III.D and IV.B.2.

In Section V, we examine the measurement process more carefully. For
the bipartite Leggett-Garg test, whether the measurement is performed
or not on system $A$ is determined by the duration of unitary evolution,
at the second location $B$.  By contrast, the timing of the final
irreversible (``collapse'') stage of the measurement at $B$ is
unimportant. We show by calculation that whether the collapse occurs
before or after the subsequent evolution at $A$ is irrelevant to
the violations, making only a difference of order $\mathcal{\hbar}e^{-|\alpha|^{2}}$
to the final probabilities. Here we point out that although the results
of this paper are consistent with weak macroscopic realism (wMR),
it is possible to establish inconsistencies with quantum mechanics,
similar to an EPR-type paradox. Finally, a conclusion is given
in Section VI.

\section{Macroscopic Bell and bipartite Leggett-Garg inequalities}

We first outline and extend the Bell and bipartite Leggett-Garg inequalities
for a macroscopic system, as proposed in Refs. \cite{weak-hybrid,macro-bell-lg}.
In order to test macroscopic realism, we combine aspects of the original
Bell and Leggett-Garg proposals. As with Bell's proposal, there are
two spatially separated systems $A$ and $B$ upon which local measurements
are made. Similar to the Leggett-Garg proposal, the systems at each
location evolve dynamically according to a local Hamiltonian $H^{(A)}$
(or $H^{(B)}$), so that measurements can be made at different times.
The measurements at each site $A$ and $B$ give only two possible
outcomes, corresponding to two macroscopically distinguishable states.
The outcomes are designated as pseudo-spins, $S_{A}=\pm1$ and $S_{B}=\pm1$,
respectively.

In fact, we present two types of test of macroscopic realism. The
first (Section II.A) is a macroscopic Bell test, where for a given
run of the experiment, one makes single measurements at each site,
$A$ and $B$. For each measurement, there is a choice between two
measurement settings, corresponding to a choice between two different
times of interaction with a measurement device. The second type of
test of macroscopic realism (Section II.B) is similar to that proposed
by Leggett and Garg, except there are two sites. In a given run of
the experiment, one makes sequential measurements at different times
$t_{1}<t_{2}<t_{3}$, on the systems at the separated sites, $A$
and $B$.
\begin{figure}[t]
\begin{centering}
\includegraphics[width=1\columnwidth]{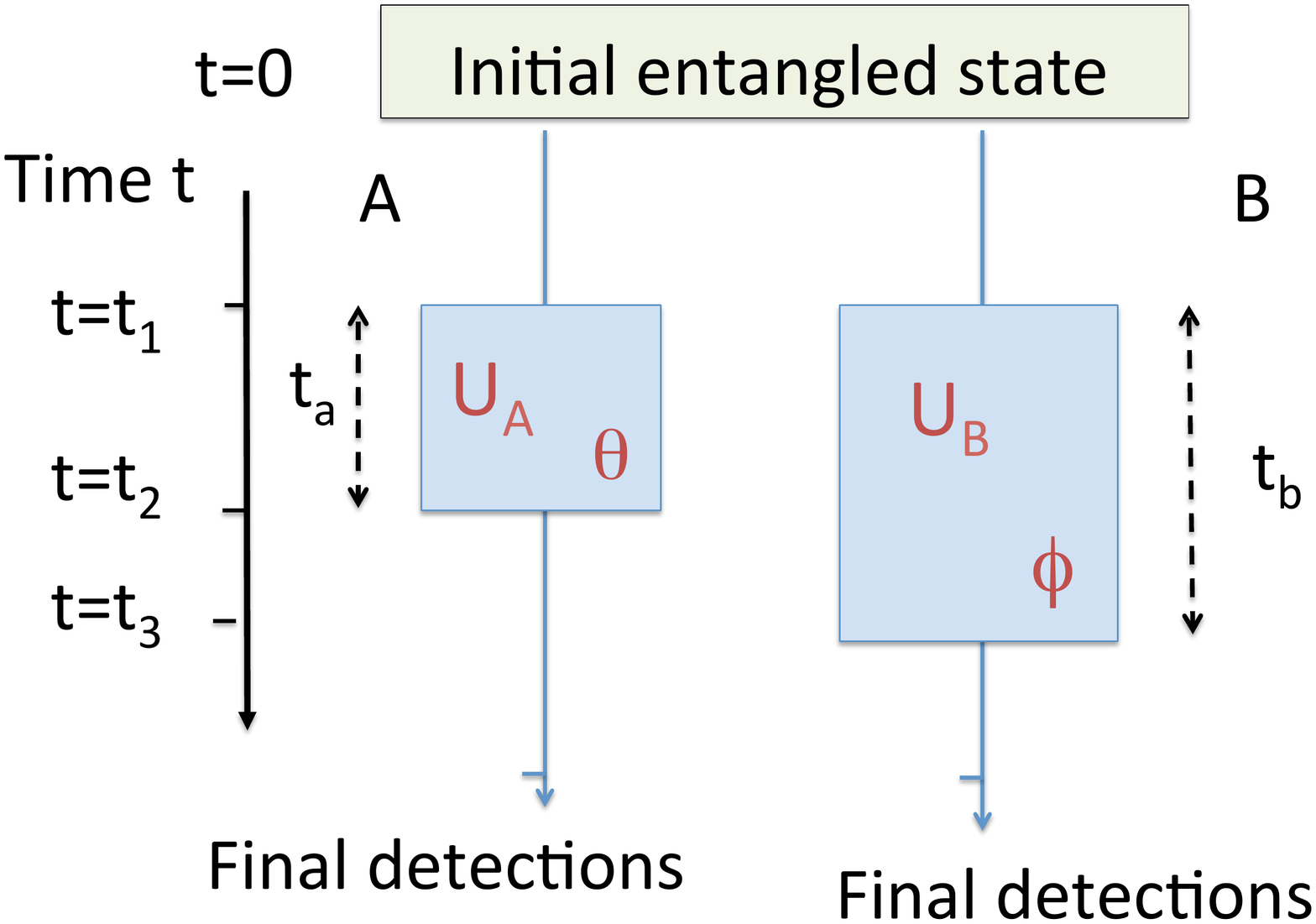}
\par\end{centering}
\caption{Bell test of macroscopic local realism using local dynamics: Measurements
$\hat{M}_{\theta}^{(A)}$ and $\hat{M}_{\phi}^{(B)}$ are made at
the sites $A$ and $B$ respectively. The measurements have two stages.
At each site, there is a choice of two measurement settings ($\theta$
and $\theta'$, and $\phi$ and $\phi'$) corresponding to times $t_{a}$
and $t'_{a}$ at $A$, and $t_{b}$ and $t'_{b}$ at $B$, during
which the system interacts with a local medium. The interactions at
$A$ and $B$ correspond to unitary transformations $U_{A}(\theta)$
and $U_{B}(\phi)$ respectively. After the unitary interactions, final
pointer measurements $\hat{M}^{(A)}$ and $\hat{M}^{(B)}$ are made
each giving two macroscopically distinct outcomes (denoted $\pm1$).}
\end{figure}

\subsection{Macroscopic Bell test}

\subsubsection{Deterministic macroscopic (local) realism}

We consider the situation of Figure 1. One prepares the state of the
system at a time $t=0$. At each site there are two possible measurement
settings ($\theta$ and $\theta'$ at $A$, and $\phi$ and $\phi'$
at $B$) and for each of these measurements there are only two outcomes,
corresponding to states of the system that are macroscopically distinct.
We refer to the outcomes $S_{A}=\pm1$ and $S_{B}=\pm1$ as ``spin''
outcomes, and label the measurements at sites $A$ and $B$ as $\hat{M}_{\theta}^{(A)}$
and $\hat{M}_{\phi}^{(B)}$ respectively. Distinguishing only between
macroscopically distinct outcomes, the measurements $\hat{M}_{\theta}^{(A)}$
and $\hat{M}_{\phi}^{(B)}$ can allow a significant error, beyond
the level of $\hbar,$ as we will show by example in Sections III
and IV. As such, we refer to the measurements as coarse-grained, or
macroscopic, measurements.

The assumption of \emph{macroscopic realism} (MR), as specified by
Leggett and Garg \cite{legggarg-1}, asserts that each system prior
to the measurement $\hat{M}_{\theta}^{(A)}$ and $\hat{M}_{\phi}^{(B)}$
is in one or other of the two macroscopically distinguishable states,
which have a definite outcome for the spin. Here, we do not wish to
restrict that the two macroscopically distinct states are \emph{specific}
states, $\psi_{1}$ and $\psi_{2}$, or even that they are quantum
states. The definition of MR is generalised to postulate that prior
to the measurement $\hat{M}_{\theta}^{(A)}$ and $\hat{M}_{\phi}^{(B)}$,
the system is in a state for which the outcome of the macroscopic,
coarse-grained measurement $\hat{M}_{\theta}^{(A)}$ or $\hat{M}_{\phi}^{(B)}$
is definite, being either $+1$ or $-1$. Assuming MR, macroscopic
hidden variables $\lambda_{\theta}^{(A)}$ and $\lambda_{\phi}^{(B)}$
with values $+1$ or $-1$ can be ascribed to the systems $A$ and
$B$ at the time immediately prior to the measurement, the values
giving the predetermined outcomes for the measurements $\hat{M}_{\theta}^{(A)}$
and $\hat{M}_{\phi}^{(B)}$ respectively. These variables, introduced
by Leggett and Garg, are hidden variables, since they are not explicitly
given by quantum mechanics.

If there is a choice of measurement settings $\theta$ and $\theta'$
at a given site, then one assumes (based on the assumption of MR)
that \emph{two} hidden variables $\lambda_{\theta}^{(A)}$ and $\lambda_{\theta'}^{(A)}$
describe the state of the system as it exists prior to the time the
choice is made, to predetermine the results of \emph{both} measurements,
$\hat{M}_{\theta}^{(A)}$ and $\hat{M}_{\theta'}^{(A)}$. Similar
to the original Bell derivation \cite{Bell-2,det-bell}, we call the
definition of MR as given here \emph{deterministic macroscopic realism
}(dMR)\emph{.}

Assuming locality, that the value of the macroscopic hidden variable
$\lambda$ for the system at one site cannot be changed by the local
measurement at the other site \cite{bell-brunner-rmp,CShim-review,Bell-2,det-bell},
it is straightforward to derive the Clauser-Horne-Shimony-Holt (CHSH)
Bell inequality, $|B|\leq2$, where \cite{CShim-review}
\begin{eqnarray}
B & = & E(\theta,\phi)-E(\theta,\phi')+E(\theta',\phi)+E(\theta',\phi')\label{eq:chsh}
\end{eqnarray}
Here $\theta$ and $\theta'$ ($\phi$ and $\phi'$) are two measurement
settings at sites A (and B) respectively, and $E(\theta,\phi)$ is
the expectation value for the product $S_{A}S_{B}$, for the choice
of settings, $\hat{M}_{\theta}^{(A)}$ and $\hat{M}_{\phi}^{(B)}$.
The combined assumptions of locality and deterministic macroscopic
realism are referred to as \emph{deterministic macroscopic local realism
}(MLR)\emph{, }or simply\emph{ deterministic macroscopic realism,}
as the condition of locality is naturally implied by dMR for two sites
\cite{det-bell}. The derivation of (\ref{eq:chsh}) follows from
the assumption of MLR, which implies $E(\theta,\phi)=\langle S_{A}S_{B}\rangle=\langle\lambda_{\theta}^{(A)}\lambda_{\phi}^{(B)}\rangle$.

The local measurements $\hat{M}_{\theta}^{(A)}$ and $\hat{M}_{\phi}^{(B)}$
have two stages: a unitary stage $U$ for which there is an interaction
with an analyzer device, and a final stage $\hat{M}$ which includes
an irreversible coupling to a reservoir i.e. to a detector. The unitary
stage determines the measurement setting, and is analogous to the
transmission of a photon or spin-$1/2$ particle through a polarizing
beam splitter or Stern-Gerlach analyzer. Thus, in the traditional
Bell tests, $\theta$ and $\phi$ are polarizer or analyzer angles,
determining the direction of polarisation or spin that is to be measured
at each site. The final pointer stage $\hat{M}$ corresponds to detection
in one of the arms of the polarizing beam splitter.

For the examples given in this paper, following Refs. \cite{macro-bell-lg,manushan-cat-lg},
the unitary stage of the measurement at each site comprises an interaction
with a nonlinear medium, and the measurement setting is determined
by the time duration of the interaction. This need not be a choice
of time itself, but can instead be a choice of the degree of nonlinearity
of the medium, the choice of measurement setting being made by a switch.
We identify at each site two choices of times for the measurements:
$t_{a}$ and $t_{a}'$ ($t_{b}$ and $t_{b}'$) at $A$ (and $B$)
respectively. The Bell inequality (\ref{eq:chsh}) becomes $|B|\leq2$
where 
\begin{eqnarray}
B & = & E(t_{a},t_{b})-E(t'_{a},t'_{b})+E(t'_{a},t_{b})+E(t'_{a},t'_{b})\label{eq:chshtime}
\end{eqnarray}
Modelling the measurement in this way as an interaction with a device
emphasises that the measurement will occur over a finite time interval.
To justify the locality assumption, it is assumed that the measurement
events at each site are spacelike separated.

\subsubsection{ Macroscopic local causality}

When deriving (\ref{eq:chsh}) and (\ref{eq:chshtime}), it is assumed
that\emph{ two} macroscopic hidden variables describe each system
at the time immediately prior to the measurement ($t=0$) \cite{bell-brunner-rmp,CShim-review,Bell-2,det-bell}.
These are $\lambda_{\theta}^{(A)}$ and $\lambda_{\theta'}^{(A)}$
for system $A$, and $\lambda_{\phi}^{(B)}$ and $\lambda_{\phi'}^{(B)}$
for system $B$. Each of these variables assumes values $+1$ or $-1$.
Thus, the system at $A$ is in a state with simultaneous predetermined
values for the measurements $\hat{M}_{\theta}$ and $\hat{M}_{\theta'}$,
and similarly for $B$.

However, it is well known that the Bell inequality (\ref{eq:chsh})
can be derived with a weaker assumption allowing for a stochastic
interaction with a local measurement device \cite{det-bell,CShim-review}.
Here, one assumes that the system at time $t=0$ is in a hidden variable
state given by a set of hidden variables $\{\lambda\}$ with probability
density $\rho(\lambda)$. One defines the probability $P_{A}(\pm1|\lambda,\theta)$
for obtaining a spin outcome $+1$ or $-1$ at site $A$, given the
system prior to measurement is in the state $\{\lambda\}$ and given
the choice of measurement setting $\theta$ for the local device at
$A$. The probability $P_{B}(\pm1|\lambda,\phi)$ is defined similarly,
for the spin outcome $\pm1$ at $B$, given the measurement setting
$\phi$ for the local measurement device at $B$. The locality assumption
is that probability $P_{A}(\pm1|\lambda,\theta)$ does not depend
on $\phi$, and $P_{B}(\pm1|\lambda,\phi)$ does not depend on $\theta$.
These assumptions can be viewed as a single assumption, local causality
\cite{det-bell}.

Thus, the Bell inequalities (\ref{eq:chsh}) and (\ref{eq:chshtime})
follow from a weaker assumption, that we refer to as \emph{macroscopic
local causality} (MLC). The assumption is similar to local causality,
except here one postulates that prior to the measurement, the system
is in a hidden variable state which gives a definite probability for
just two macroscopically distinct outcomes. In this paper, the Bell
inequalities can be derived from either MLR or MLC, and we will use
MLR to imply either definition.

\subsection{Bipartite Leggett-Garg tests}

The Leggett-Garg inequalities were derived for the situation where
one considers succcessive measurements on a single system, at times
$t_{i}$. Here, following \cite{macro-bell-lg,weak-hybrid} and different
to the original treatment \cite{legggarg-1}, we consider two systems
$A$ and $B$ (Figure 2). Measurements $\hat{M_{i}}^{(A)}$ and $\hat{M}_{j}^{(B)}$
are made on systems $A$ and $B$ at times $t_{i}$ and $t_{j}$ respectively.
The measurements give two outcomes, $+1$ and $-1$, corresponding
to two macroscopically distinguishable states. However, here there
is no choice of measurement setting at a given time: Rather, the unitary
rotation stage $U_{A}$ (or $U_{B}$) of the measurement is considered
to be part of the dynamics. In the examples we consider in this paper,
the measurements $\hat{M_{i}}^{(A)}$ and $\hat{M}_{j}^{(B)}$ that
occur at the times $t_{i}$ and $t_{j}$ are analogous to the pointer
measurements, $\hat{M}^{(A)}$ and $\hat{M}^{(B)}$, of Figure 1.

The Leggett-Garg inequalities are derived based on the assumption
of macroscopic realism \cite{legggarg-1}.  Hence one may identify
specific macroscopic hidden variables $\lambda_{i}^{(A)}$ and $\lambda_{i}^{(B)}$
to describe the states immediately prior to the pointer measurements
$\hat{M_{i}^{(A)}}$ and $\hat{M}_{j}^{(B)}$. Different to the Bell
derivation of Section II.A.1 however, one does not ascribe to the
system simultaneously at a time $t_{i}$ predetermined results $\lambda_{i}^{(A)}$
and $\lambda_{j}^{(A)}$ for \emph{both} future measurements, $\hat{M}_{i}^{(A)}$
and $\hat{M}_{j}^{(A)}$. What was the local unitary interaction
($U_{A}$ or $U_{B}$) with the measurement device in the Bell test
is now the local trajectory. One therefore assumes macroscopic realism
for a\emph{ }measurement with just a single setting, at the time $t_{i}$.
This means that the Leggett-Garg-Bell inequalities for the systems
of interest may be derived based on the assumption of \emph{weak macroscopic
realism} (wMR): that for a system $A$ prepared in a superposition
of macroscopically distinct pointer states, the system can be described
by a single hidden variable $\lambda_{i}$ at that time $t_{i}$ (and
this is not affected by a space-like separated measurement event at
the different site $B$).

The traditional Leggett-Garg test invokes a second assumption, that
it is in principle possible to determine which of the two macrosopically
distinct states the system is in at any given time, by performing
a \emph{noninvasive measurement} that has a negligible effect on the
subsequent dynamics \cite{legggarg-1}. This assumption is necessary
because the traditional Leggett-Garg tests involve measurements at
different times performed on the same system, and there is the possibility
of direct disturbance of the system due to measurement. In this paper,
we follow \cite{macro-bell-lg}, and consider two systems (Figure
2). One assumes validity of (weak) macroscopic realism, but the additional
assumption of noninvasive measureability is replaced by the assumption
of locality. Specifically, for measurements at $A$ and $B$ that
are spacelike separated events, one assumes the hidden variable value
$\lambda_{i}^{(A)}$ is not changed by whether the measurement at
$B$ takes place or not, and similarly for $\lambda_{j}^{(B)}$.

\begin{figure}[t]
\begin{centering}
\includegraphics[width=1\columnwidth]{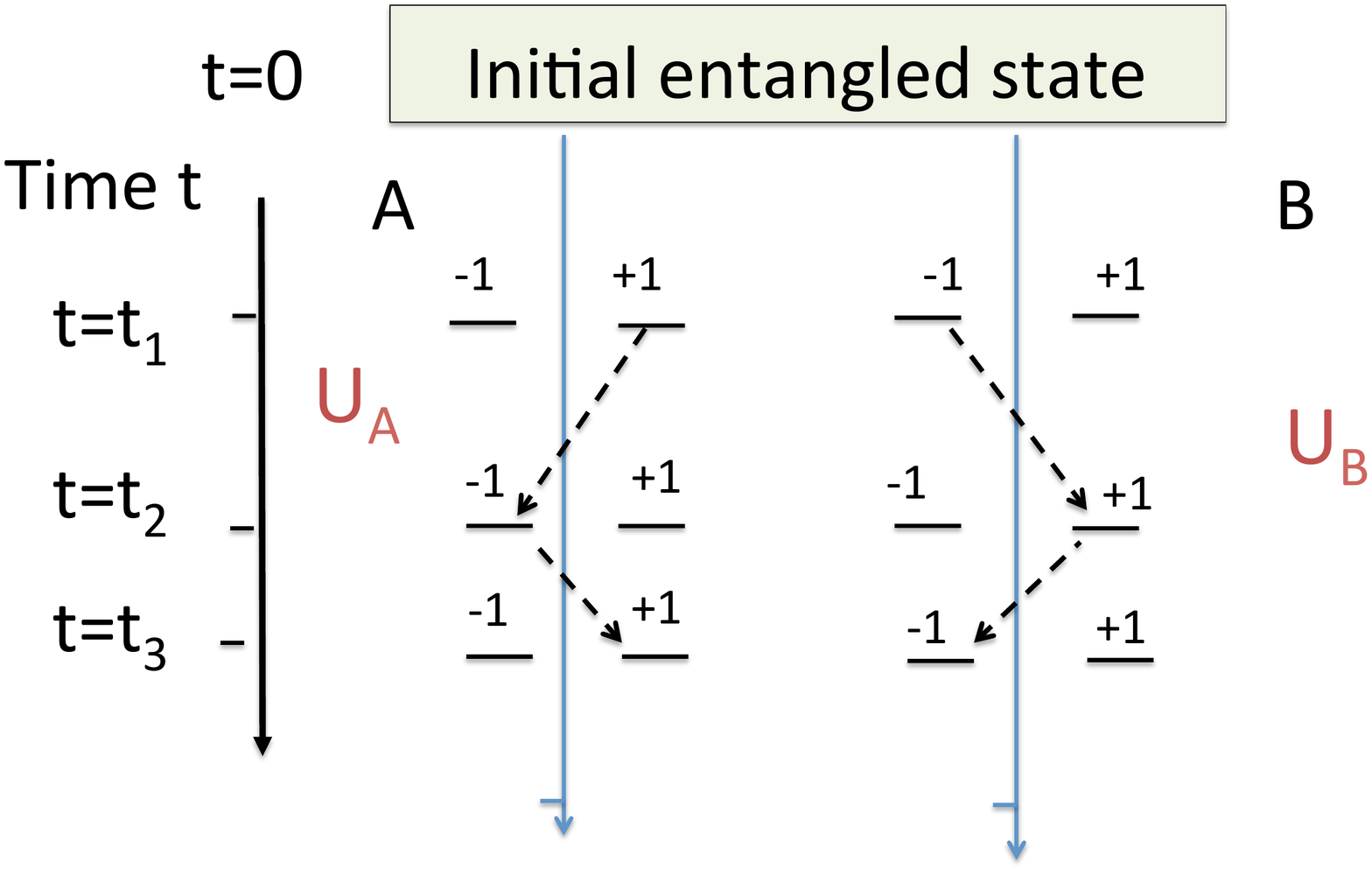}\vspace{-1cm}
\par\end{centering}
\caption{A Leggett-Garg test of macrorealism using bipartite systems. Two systems
$A$ and $B$ prepared in an entangled state at time $t_{1}$ undergo
dynamical evolution according to local unitary operators, $U_{A}(t)$
and $U_{B}(t)$. At the later times $t_{2}$ and $t_{3}$, each system
is found to be in one or other of two macroscopically distinct states
$-1$ or $+1$, and the outcomes are anti-correlated. One can infer
the outcome for $A$ by measuring at $B$. A possible evolution is
shown by the dashed lines.}
\end{figure}

We consider measurements made at times $t_{1}$ and $t_{3}$ at site
$A$, and at times $t_{2}$ and $t_{4}$ at site $B$, where $t_{1}<t_{2}<t_{3}<t_{4}$.
Let us denote the spin outcome $S_{A}$ at $A$ for the measurement
made at time $t_{i}$ by $S_{i}^{(A)}$. The spin $S_{j}^{(B)}$ is
defined similarly, for site $B$. With the assumptions of macroscopic
realism and noninvasive measurability (locality), the Leggett-Garg-Bell
inequality
\begin{eqnarray}
-2 & \leq & E(t_{1},t_{2})-E(t_{1},t_{4})+E(t_{2},t_{3})+E(t_{3},t_{4})\leq2\nonumber \\
\label{eq:lg-double}
\end{eqnarray}
where $E(t_{i},t_{j})=\langle S_{i}^{(A)}S_{j}^{(B)}\rangle$ can
be shown to hold \cite{legggarg-1}. This inequality was derived as
a Leggett-Garg inequality in Ref. \cite{legggarg-1}, and as a Leggett-Garg-Bell
inequality, in Refs. \cite{macro-bell-lg,weak-hybrid}. Since the
expectation values $E(t_{i},t_{j})$ of the inequality involve measurements
made at diffferent sites, the inequality follows based on the assumption
of locality, which justifies that $E(t_{i},t_{j})=\langle S_{i}^{(A)}S_{j}^{(B)}\rangle=\langle\lambda_{i}^{(A)}\lambda_{j}^{(B)}\rangle$.
We note that while we assume the time order $t_{1}<t_{2}<t_{3}<t_{4}$,
in fact because we have two localised sites, we need only take $t_{1}<t_{3}$
and $t_{2}<t_{4}$.

A three-time Leggett-Garg-Bell inequality
\begin{eqnarray}
-3 & \leq & E(t_{1},t_{2})-E(t_{1},t_{3})+E(t_{2},t_{3})\leq1,\label{eq:lg-double-three-1}
\end{eqnarray}
where $E(t_{i},t_{j})=\langle S_{i}^{(B)}S_{j}^{(A)}\rangle$ is also
useful. This is based on the original three-time Leggett-Garg inequality,
derived for the single-system set-up, in Refs. \cite{weak-solid-state-qubits,jordan_kickedqndlg2-1}.
Here, one assumes however that a measurement at time $t_{1}$ at site
$A$ does not affect the outcome at the later time $t_{3}$ at the
same site, which raises the issue of a disturbance due to the measurement
at $t_{1}$. However, for correlated initial states, one may use that
the result for $S_{1}^{(B)}$ can be inferred from a measurement of
$S_{1}^{(A)}$. Thus, assuming locality, one may deduce \cite{macro-bell-lg}:
\begin{equation}
E(t_{1},t_{3})=\langle S_{1}^{(A)}S_{3}^{(A)}\rangle=-\langle S_{1}^{(B)}S_{3}^{(A)}\rangle\label{eq:equate}
\end{equation}

We note that the no-signalling-in-time (NSIT) equality has been shown
to give a stronger test of Leggett-Garg's macrorealism \cite{NSTmunro-1,nst}.
However, it is more challenging to use this equality for a bipartite
test.

\subsection{Summary of assumptions}

Since we are to show violation of the macroscopic Bell and Leggett-Garg-Bell
inequalities, it is important to summarise what is the weakest (i.e.
minimal\emph{)} assumption needed to derive the inequalities. The
Bell inequalities (\ref{eq:chshtime}) can be derived from the minimal
assumption of macroscopic local causality (MLC), as in Section 1.A.2.
In this paper, however, we use the terms macroscopic local realism
(MLR) and macroscopic local causality (MLC) interchangeably. The Leggett-Garg-Bell
inequalities (\ref{eq:lg-double}) and (\ref{eq:lg-double-three-1})
on the other hand are derived assuming \emph{weak macroscopic realism},
combined with macroscopic noninvasiveness (NIM) of the measurement,
which for two sites is justified by macroscopic locality (ML).

\section{Violating the macroscopic Bell inequalities}

We consider the system of Figure 1 prepared in the initial ``cat''
state \cite{cat-bell-wang-1}
\begin{equation}
|\psi_{Bell}\rangle=\mathcal{N}(|\alpha\rangle_{a}|-\beta\rangle_{b}-|-\alpha\rangle_{a}|\beta\rangle_{b})\label{eq:cat-1}
\end{equation}
Here $|\alpha\rangle$ and $|\beta\rangle$ are coherent states
for two modes, labelled $a$ and $b$, and for convenience we assume
$\alpha$, $\beta$ to be real and positive. We will ultimately take
$\alpha$, $\beta\rightarrow\infty$. The normalisation constant is
$\mathcal{N}=\frac{1}{\sqrt{2}}\{1-\exp(-2\left|\alpha\right|^{2}-2\left|\beta\right|^{2})\}^{-1/2}$.
In the limit where $\alpha$, $\beta$ are large, we note that $|\alpha\rangle_{a}$
and $|-\alpha\rangle_{a}$ become orthogonal, and similarly $|\beta\rangle_{b}$
and $|-\beta\rangle_{b}$, so that $\mathcal{N}\rightarrow\frac{1}{\sqrt{2}}$.
In this limit, we may write the cat state as a macroscopic two-qubit
state
\begin{equation}
|\psi_{Bell}\rangle=\frac{1}{\sqrt{2}}(|+\rangle_{a}|-\rangle_{b}-|-\rangle_{a}|+\rangle_{b})\label{eq:cat-1-2}
\end{equation}
Here $|+\rangle_{a}$ and $|-\rangle_{a}$ are orthogonal states giving
an outcome of $+1$ and $-1$ respectively, for a measurement $\hat{M}^{(A)}$
of the sign $S_{A}$ of the coherent amplitude, $\alpha$. Similarly,
$|+\rangle_{b}$ and $|-\rangle_{b}$ are orthogonal states, with
outcome $+1$ and $-1$ for the measurement $\hat{M}^{(B)}$ of the
sign $S_{B}$ of the amplitude $\beta$. The two modes $a$ and $b$
are spatially separated, being located at sites $A$ and $B$, respectively.
The sign $S$ of the coherent amplitude is taken to be the value of
the ``spin'' for each mode (site).

\subsection{Bell violations using unitary local transformations}

At locations $A$ and $B$, we assume local unitary transformations
$U_{A}(t_{a})$ and $U_{B}(t_{b})$ take place, as in Figure 1. States
$|\pm\rangle_{a}$ and $|\pm\rangle_{b}$ are transformed after a
time $t_{a}$ and $t_{b}$ according to (at site $A$)
\begin{eqnarray}
U_{A}(t{}_{a})|+\rangle_{a} & = & a|+\rangle_{a}+i\sqrt{1-a^{2}}|-\rangle_{a}\nonumber \\
U_{A}(t{}_{a})|-\rangle_{a} & = & a|-\rangle_{a}+i\sqrt{1-a^{2}}|+\rangle_{a}\label{eq:trans1-2}
\end{eqnarray}
and (at site $B$)
\begin{eqnarray}
U_{B}(t{}_{b})|+\rangle_{b} & = & b|+\rangle_{b}+i\sqrt{1-b^{2}}|-\rangle_{b}\nonumber \\
U_{B}(t{}_{b})|-\rangle_{b} & = & b|-\rangle_{b}+i\sqrt{1-b^{2}}|+\rangle_{b}\label{eq:trans2-2}
\end{eqnarray}
For a system prepared in the superposition $|\psi_{Bell}\rangle$
of eqn. (\ref{eq:cat-1-2}), the system will evolve into the final
state\textcolor{blue}{\emph{}}
\begin{align}
|\psi_{f}\rangle & =\frac{1}{\sqrt{2}}(U_{A}|+\rangle_{a}U_{B}|-\rangle_{b}-U_{A}|-\rangle_{a}U_{B}|+\rangle_{b})\nonumber \\
 & =\frac{1}{\sqrt{2}}\{(ab+\bar{a}\bar{b})(|+\rangle_{a}|-\rangle_{b}-|-\rangle_{a}|+\rangle_{b})\nonumber \\
 & \thinspace\thinspace{\color{blue}{\normalcolor +}}i(\bar{a}b-a\bar{b})(|+\rangle_{a}|+\rangle_{b}-|-\rangle_{a}|-\rangle_{b})\}\label{eq:finalstate}
\end{align}
where we have abbreviated $\bar{a}=\sqrt{1-a^{2}}$ and $\bar{b}=\sqrt{1-b^{2}}.$
Using that the states $|\pm\rangle$ are orthogonal, and defining
the measurement $\hat{S}$ which measures the sign $S$ (which we
label as its ``spin''), according to $\hat{S}^{(A)}|\pm\rangle_{a}=\pm|\pm\rangle_{a}$
and $\hat{S}^{(B)}|\pm\rangle_{b}=\pm|\pm\rangle_{b}$, we find that
the expectation value $E(t_{a},t_{b})$ of the spin product $\hat{S}^{(A)}\hat{S}^{(B)}$
is
\begin{equation}
E(t_{a},t_{b})=|\bar{a}b-a\bar{b}|^{2}-|ab+\bar{a}\bar{b}|^{2}\label{eq:etimes}
\end{equation}

For the traditional Bell experiment where the transformation is achieved
using a polarising beam splitter, we have $a=\cos\theta$ and $b=\cos\phi$,
and
\begin{equation}
E(\theta,\phi)=-\cos2(\theta-\phi)\label{eq:Ebell}
\end{equation}
It is known that the choice of angles $\theta=0$, $\phi=\pi/8$,
$\theta'=\pi/4$ and $\phi'=3\pi/8$ in the Bell inequality (\ref{eq:chsh})
will lead to $B=-2\sqrt{2}$ which violates the inequality.

\subsection{Unitary transformations via nonlinear media}

Now we consider how to realise the transformations (\ref{eq:trans1-2})
and (\ref{eq:trans2-2}). We seek a unitary transformation $U_{\theta}^{(A)}\equiv U_{A}(\theta)$
such that at site $A$ the transformation
\begin{eqnarray}
U_{A}(\theta)|\alpha\rangle & = & \cos\theta|\alpha\rangle+i\sin\theta|-\alpha\rangle\nonumber \\
U_{A}(\theta)|-\alpha\rangle & = & \cos\theta|-\alpha\rangle+i\sin\theta|\alpha\rangle\label{eq:rot}
\end{eqnarray}
holds, for the particular angle choices $\theta=0$ and $\theta=\pi/4$.
At site $B$, we seek a transformation such that
\begin{eqnarray}
U_{B}(\phi)|\beta\rangle & = & \cos\phi|\beta\rangle+i\sin\phi|-\beta\rangle\nonumber \\
U_{B}(\phi)|-\beta\rangle & = & \cos\phi|-\beta\rangle+i\sin\phi|\beta\rangle\label{eq:rot-1}
\end{eqnarray}
for the angle choices $\phi=\pi/8$, and $\phi=3\pi/8$. We drop subscripts
$a$ and $b$, where the meaning is clear.

These transformations can be achieved by interacting each single mode
system with a nonlinear medium at the given site, where the interaction
Hamiltonian is \textcolor{black}{$H_{NL}=\Omega\hat{n}^{k}$ }($k>2$
and $k$ even)\textcolor{black}{{} \cite{manushan-cat-lg,yurke-stoler-1,wrigth-walls-gar-1,macro-bell-lg}.
Here  $\hat{n}$ is the mode number operator and $\Omega$ is the
nonlinear coefficient. We choose units such that $\hbar=1$. The
interaction takes place independently for each mode $a$ and $b$,
and we denote the respective Hamiltonians as $H_{NL}^{(A)}$ and $H_{NL}^{(B)}$.}
Thus we define local interactions at each site
\begin{equation}
H_{NL}^{(A)}=\Omega\hat{n}_{a}^{k},\ \ H_{Nl}^{(B)}=\Omega\hat{n}_{b}^{k}\label{eq:ham_NL}
\end{equation}
\textcolor{black}{The boson destruction mode operators for modes $a$
and $b$ are denoted by $\hat{a}$ and $\hat{b}$, respectively, and
the number operators are $\hat{n}_{a}=\hat{a}^{\dagger}\hat{a}$ and
$\hat{n}_{b}=\hat{b}^{\dagger}\hat{b}$.}
\begin{figure}[t]
\begin{centering}
\includegraphics[width=1\columnwidth]{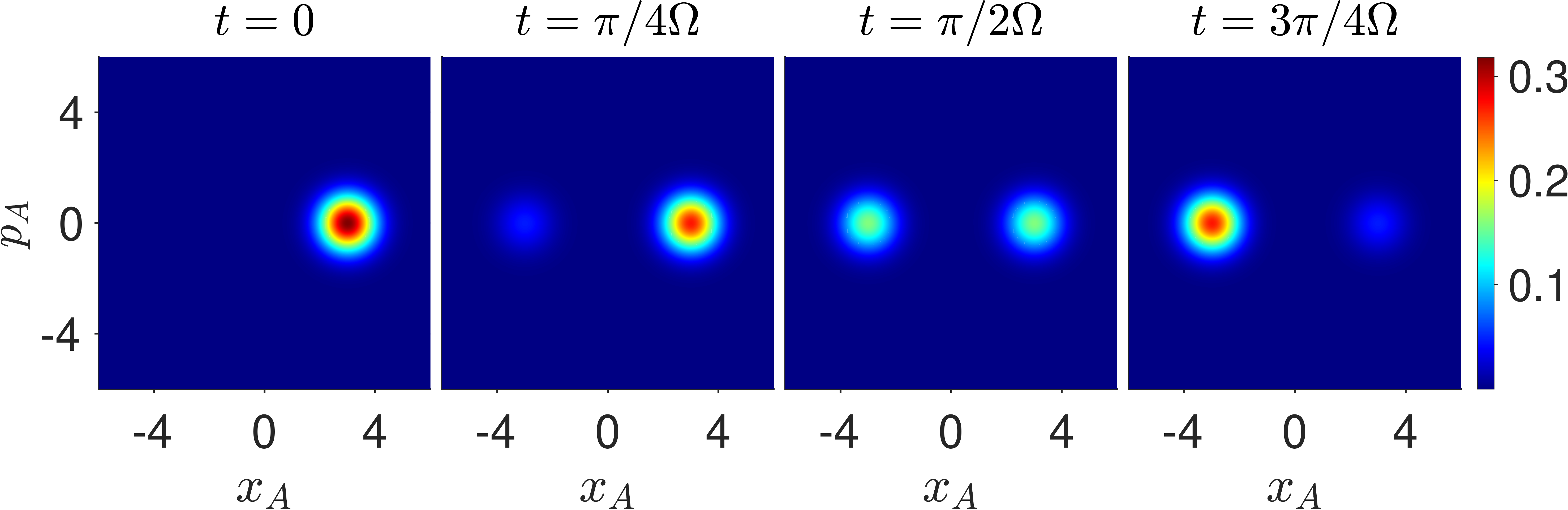}
\par\end{centering}
\begin{centering}
\par\end{centering}
\caption{An initial coherent state $|\alpha\rangle$ evolving according to
the Hamltonian $H_{NL}^{(A)}$, $k=4$, $\alpha=3$. The Figure shows
contour plots of the Q function $Q(x_{A},p_{A})$\textcolor{red}{{}
}for the states created at a time $t$. For the times shown, a
superposition of the states $|-\alpha\rangle$ and $|\alpha\rangle$
is formed.\textcolor{red}{}}
\end{figure}

\begin{figure}[t]
\begin{centering}
\par\end{centering}
\begin{centering}
\includegraphics[width=1\columnwidth]{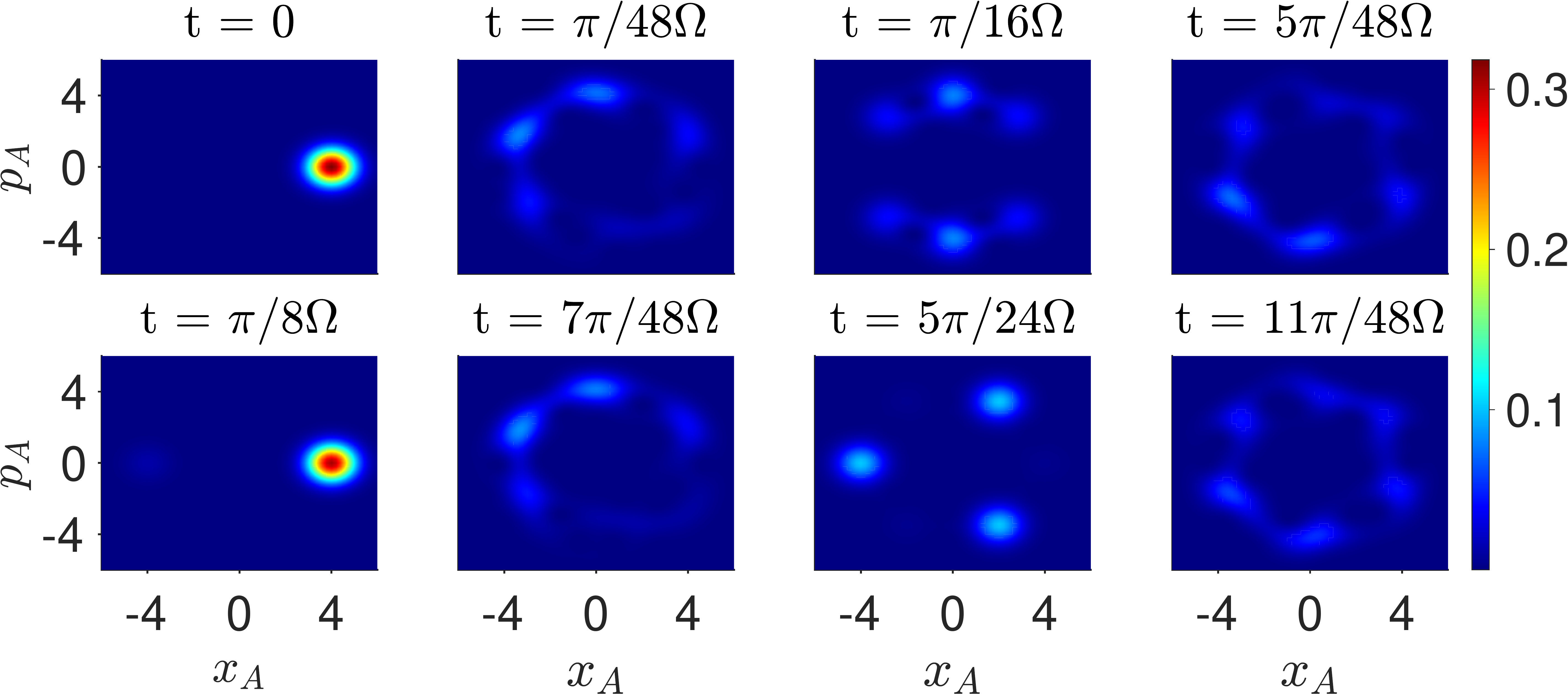}
\par\end{centering}
\begin{centering}
\par\end{centering}
\caption{The detailed dynamics of the creation of the superposition state shown
in Figure 4. The system is not a superposition of states $|-\alpha\rangle$
and $|\alpha\rangle$ at all times $t$. Here, $\alpha=4$.\textcolor{red}{}}
\end{figure}

\textcolor{black}{One may solve for the dynamics given by $H_{NL}$.
If the system $A$ is prepared in a coherent state $|\alpha\rangle_{a}$,
the state after time $t$ is \cite{yurke-stoler-1}
\begin{eqnarray}
\left|\alpha,t_{a}\right\rangle  & = & U_{A}(t_{a})|\alpha\rangle_{a}\nonumber \\
 & = & \exp[-\frac{\left|\alpha\right|^{2}}{2}]\underset{n=0}{\sum^{\infty}}\frac{\alpha^{n}\exp(-i\phi_{n})}{\sqrt{n!}}\left|n\right\rangle _{a}\label{eq:cat-solns-1}
\end{eqnarray}
where $U_{A}(t_{a})=e^{-iH_{NL}^{(A)}t_{a}/\hbar}$ and $\phi_{n}=\varOmega t_{a}n^{k}$.
}\textcolor{black}{For an initial coherent state $|\beta\rangle_{b}$
at $B$, the state after a time $t$ is given similarly, as $\left|\beta,t_{b}\right\rangle =U_{B}(t_{b})|\beta\rangle_{b}$
where $U_{B}(t_{b})=e^{-iH_{NL}^{(B)}t_{b}/\hbar}$. These solutions
are well known \cite{manushan-cat-lg,yurke-stoler-1} and are depicted
graphically in Figures 3 and 4 in terms of the $Q$ function, defined}
as $Q(x_{A},p_{A})=\frac{1}{\pi}\langle\alpha_{0}|\rho|\alpha_{0}\rangle$
\cite{Husimi-Q-1}. Here $\alpha_{0}=x_{A}+ip_{A}$ and $|\alpha_{0}\rangle$
is a coherent state\textcolor{black}{. At certain intervals of time,
the system evolves into a superposition of the macroscopically distinct
coherent states }$|-\alpha\rangle$ and $|\alpha\rangle$\textcolor{black}{{}
(Figure 4).}

Suppose at time $t_{a}=t_{b}=0$, the systems are in the initial states
$|\alpha\rangle$ and $|\beta\rangle$. After a time $t_{b}=\pi/4\Omega$,
the system at $B$ is in the state $U_{B}(\pi/4\Omega)|\beta\rangle$
(second picture of Figure 3). This gives the transformation of eq.
(\ref{eq:rot-1}) with $\phi=\pi/8$ \cite{manushan-cat-lg}. We
write the state at this time as
\begin{eqnarray}
U_{\pi/8}^{(B)}|\beta\rangle & = & e^{-i\pi/8}\{(\cos\pi/8|\beta\rangle+i\sin\pi/8|-\beta\rangle\}\nonumber \\
\label{eq:state3}
\end{eqnarray}
After a time $t_{a}=\pi/2\Omega$, the state is $U_{A}(\pi/2\Omega)|\alpha\rangle$
which corresponds to the transformation eq. (\ref{eq:rot}) with $\theta=\pi/4$.
We write the state at $t_{a}=\pi/2\Omega$ as \cite{manushan-cat-lg}
\begin{eqnarray}
U_{\pi/4}^{(A)}|\alpha\rangle & = & \frac{e^{-i\pi/4}}{\sqrt{2}}\{|\alpha\rangle+e^{i\pi/2}|-\alpha\rangle\}\label{eq:state1}
\end{eqnarray}
To obtain the transformation (\ref{eq:rot-1}) with $\phi=3\pi/8$,
we consider the state at time $t_{b}=3\pi/4\Omega$ \cite{manushan-cat-lg},
which is given by $U_{B}(3\pi/4\Omega)|\beta\rangle$. The state at
the time $t_{b}=3\pi/4\Omega$ is
\begin{eqnarray}
U_{3\pi/8}^{(B)}|\beta\rangle & = & e^{i\pi/8}\{(\cos3\pi/8|\beta\rangle+i\sin3\pi/8|-\beta\rangle\}\nonumber \\
\label{eq:state6}
\end{eqnarray}
These transformations are shown graphically in Figure 3. The transformations
at each site $A$ and $B$ are summarised as 
\begin{equation}
U_{A}(t_{a})=e^{-iH_{NL}^{(A)}t_{a}/\hbar},\ \ U_{B}(t_{b})=e^{-iH_{NL}^{(B)}t_{b}/\hbar}\label{eq:state5}
\end{equation}
where $t_{a}=0$, $t_{b}=\pi/4\Omega$, $t'_{a}=\pi/2\Omega$ and
$t_{b}'=3\pi/4\Omega$. Noting from above, $U_{B}(\pi/4\Omega)=U_{\pi/8}^{(B)}$,
$U_{A}(\pi/2\Omega)=U_{\pi/4}^{(A)}$ and $U_{B}(3\pi/4\Omega)=U_{3\pi/8}^{(B)}$.\textcolor{red}{}

\textcolor{black}{The Bell test of Section III.A requires a measurement
of the spin $\hat{S}$ for the states after the unitary rotations.
As we have seen above, the unitary parts of the measurements }$\hat{M}_{\theta}^{(A)}$
\textcolor{black}{and }$\hat{M}_{\phi}^{(B)}$\textcolor{black}{{} are
to be made by interacting with the nonlinear medium at each site.
This creates the superpositions (\ref{eq:state3}-\ref{eq:state5})
of the coherent states }$|-\alpha\rangle$ and $|\alpha\rangle$ ($|-\beta\rangle$
and $|\beta\rangle$) which are rotated relative to the initial state
$|\alpha\rangle$ ($|\beta\rangle$\textcolor{black}{). The initial
states determine the orientiation of real axis (i.e. the phase direction).
The coherent states }$|-\alpha\rangle$ and $|\alpha\rangle$ ($|-\beta\rangle$
and $|\beta\rangle$)\textcolor{black}{{} are hence the basis states
for a particular pointer measurement, $\hat{M}^{(A)}$ ($\hat{M}^{(B)}$)
that determines the sign of the coherent amplitude i.e. the value
of the macroscopic qubit. The pointer measurements are hence measurements
of the quadrature phase amplitudes \cite{yurke-stoler-1}
\begin{equation}
\hat{X}_{A}={\color{red}{\color{blue}{\color{black}\frac{1}{\sqrt{2}}}}}(\hat{a}+\hat{a}^{\dagger}),\ \ \hat{X}_{B}=\frac{1}{\sqrt{2}}(\hat{b}+\hat{b}^{\dagger})\label{eq:signs}
\end{equation}
on the evolved states at each site. The orientation of the axis that
defines quadratures is aligned with the orientation of the initial
coherent states }$|\alpha\rangle$ ($|\beta\rangle$\textcolor{black}{)
in phase space. This is set in the experiment by a local oscillator.
Finally, such a measurement involves a coupling to detectors. Let
us denote the results of the pointer measurements $\hat{X}_{A}$ and
$\hat{X}_{B}$ by $X_{A}$ and $X_{B}$ respectively. The final spin
outcome of the measurement }$\hat{M}_{\theta}^{(A)}$ \textcolor{black}{is
taken to be $S_{A}=+1$ if $X_{A}>0$, and $-1$ otherwise. Similarly,
the outcome of the measurement }$\hat{M}_{\phi}^{(B)}$ \textcolor{black}{is
$S_{B}=+1$ if $X_{B}>0$, and $-1$ otherwise.}

\subsection{Macroscopic Bell violations}

At each site $A$ and $B$, the system of Figure 1 prepared in the
cat state (\ref{eq:cat-1}) undergoes unitary transformations $U_{A}$
and $U_{B}$, given by (\ref{eq:state5}). There is the choice of
two rotation angles, corresponding to two different times, at each
site.\textcolor{black}{{} These give transformations of the form (\ref{eq:rot})
and (\ref{eq:rot-1}).}

\textcolor{black}{A violation of the Bell inequality (\ref{eq:chsh})
is obtained if we select angles }$\theta=0$ and $\theta=\pi/4$ at
site $A$, and $\phi=\pi/8$ and $\phi=3\pi/8$\textcolor{black}{{}
at site $B$ (Figure 1). Therefore, in order to violate the Bell inequality
(\ref{eq:chshtime}), we select $t_{a}=0$ (which gives rotation $\theta=0$)
and $t'_{a}=\pi/2\Omega$ (which gives rotation $\theta=\pi/4$).
Similarly at site B, we select $t_{b}=\pi/4\Omega$ (rotation $\phi=\pi/8$)
and }$t'_{b}=3\pi/(4\Omega)$ (rotation $\phi=3\pi/8$). Using the
prediction \textcolor{black}{$E(t_{a},t_{b})=E(\theta,\phi)$ given
by (\ref{eq:Ebell}) for orthogonal states undergoing the unitary
transformations (\ref{eq:trans1-2}) and (\ref{eq:trans2-2}), we
see that this choice of time-settings will give a violation of the
Bell inequality (\ref{eq:chshtime}).}
\begin{figure}[t]
\begin{centering}
\includegraphics[width=1\columnwidth]{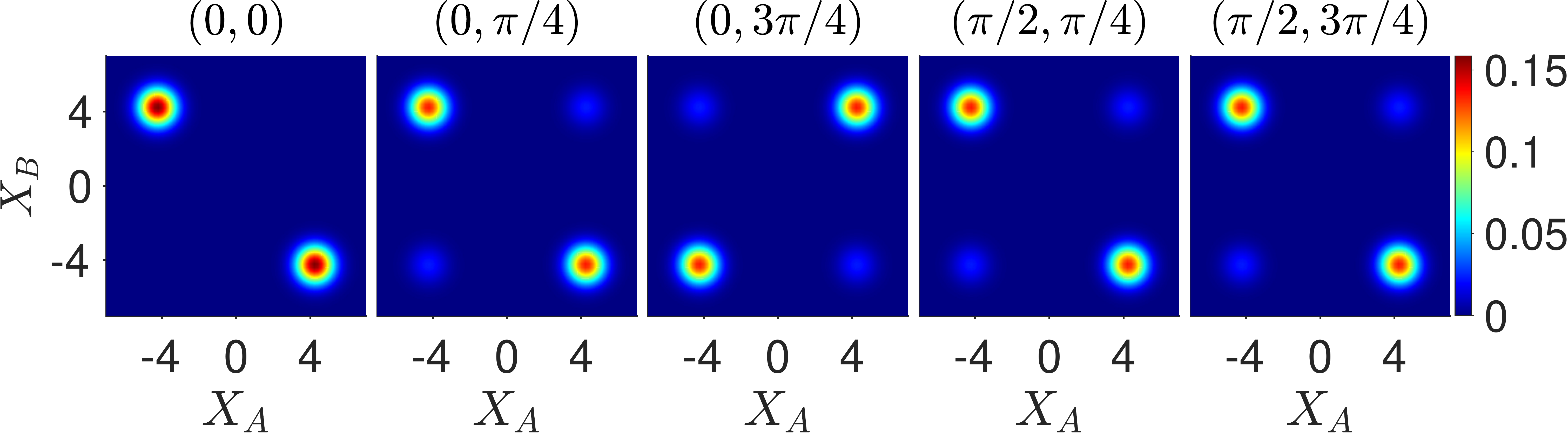}
\par\end{centering}
\begin{centering}
\par\end{centering}
\caption{Distributions giving macroscopic Bell violations: Contour plots
of $P(X_{A},X_{B})$ for the system depicted in Figure 1 with an initial
entangled state $|\psi_{Bell}\rangle$ (\ref{eq:cat-1}) and unitary
transformations $U_{A}$ and $U_{B}$ (\ref{eq:state5}), for measurement
settings $t_{a}$ and $t_{b}$. Here, $\alpha=\beta=3$ and $\Omega=1$.
For each setting $\theta$ and $\phi$ given by \emph{($t_{a}$, $t_{b}$)},
the outcomes $S_{A}$ and $S_{B}$ for $\hat{M}^{(A)}$ and $\hat{M}^{(B)}$
are $+1$ or $-1$ if the measurement of $\hat{X}_{A}$ or $\hat{X}_{B}$
yields a positive or negative value.\textcolor{red}{}}
\end{figure}

To account for the nonorthogonality due to finite amplitudes $\alpha$
and $\beta$, we evaluate the distributions $P(X_{A},X_{B})$ for
the outcomes of the joint measurement of the amplitudes $\hat{X}_{A}$
and $\hat{X}_{B}$ (eqn (\ref{eq:signs}), for each of the time-settings.
This constitutes the prediction for the final pointer measurements.
The final state after the unitary transformations is
\begin{eqnarray}
|\psi_{f}\rangle & = & \mathcal{N}\thinspace U_{A}(t_{a})U_{B}(t_{b})\Bigr(|\alpha\rangle|-\beta\rangle-|-\alpha\rangle|\beta\rangle\Bigl)\nonumber \\
\label{eq:234-1-1-1}
\end{eqnarray}
where the unitary operators are given by (\ref{eq:state5}) and $\mathcal{N}$
is the normalisation constant. The probability is
\begin{equation}
P(X_{A},X_{B})=|\langle X_{A}|\langle X_{B}|\psi_{f}\rangle|^{2}\label{eq:P}
\end{equation}
where $|X_{A}\rangle$, $|X_{B}\rangle$ are eigenstates of $X_{A}$
and $X_{B}$ respectively. Details of calculation are given in the
Appendix. Integrating gives the results needed for the evaluation
of the Bell inequality term, $B$, of eqn (\ref{eq:chshtime}). We
use $E(t_{a},t_{b})=P_{++}+P_{--}-P_{+-}-P_{-+}$ where
\begin{eqnarray}
P_{++} & = & \int_{0}^{\infty}\int_{0}^{\infty}P(X_{A},X_{B})dX_{A}dX_{B}\label{eq:p4}
\end{eqnarray}
and $P_{--}$, $P_{-+}$ and $P_{+-}$ are defined similarly for
the remaining three quadrants.

The results for the four choices of time settings are shown in Figures
5 and 6. For $\alpha,\beta\geq1.5$, the overlap of the coherent states
$|\alpha\rangle$ and $|-\alpha\rangle$ (and $|\beta\rangle$ and
$|-\beta\rangle$) is small and there is agreement with the calculations
based on the assumption of orthogonality of the states. For each time
setting, the results of the measurements \textcolor{black}{$S^{(A)}$
and $S^{(B)}$} have two possible values, $+1$ and $-1$, and there
are four joint outcomes. In Figure 5, the plots of $P(X_{A},X_{B})$
show four distinct Gaussian peaks corresponding to the probabilities
of these joint outcomes, the peaks becomes macroscopically separated
for large $\alpha$ and $\beta$. We thus verify that maximal violations
of the Bell inequality (\ref{eq:chshtime}) ($|B|=2\sqrt{2}$) are
indeed predicted for the macroscopic qubits, where measurements distinguish
macroscopically distinct states, \textcolor{black}{}as $\alpha$,
$\beta\rightarrow\infty$.
\begin{figure}[t]
\includegraphics[width=0.75\columnwidth]{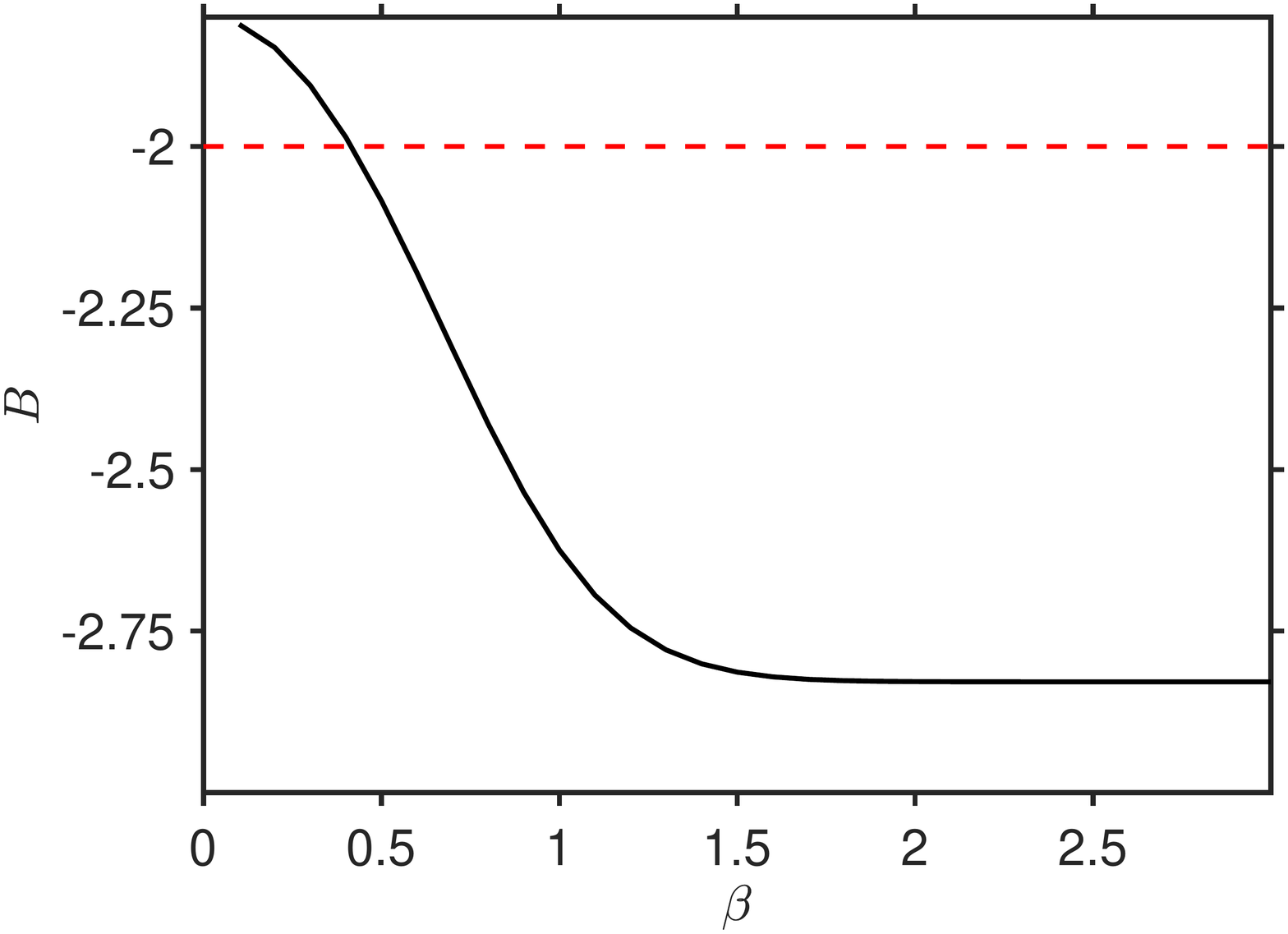}

\includegraphics[width=0.75\columnwidth]{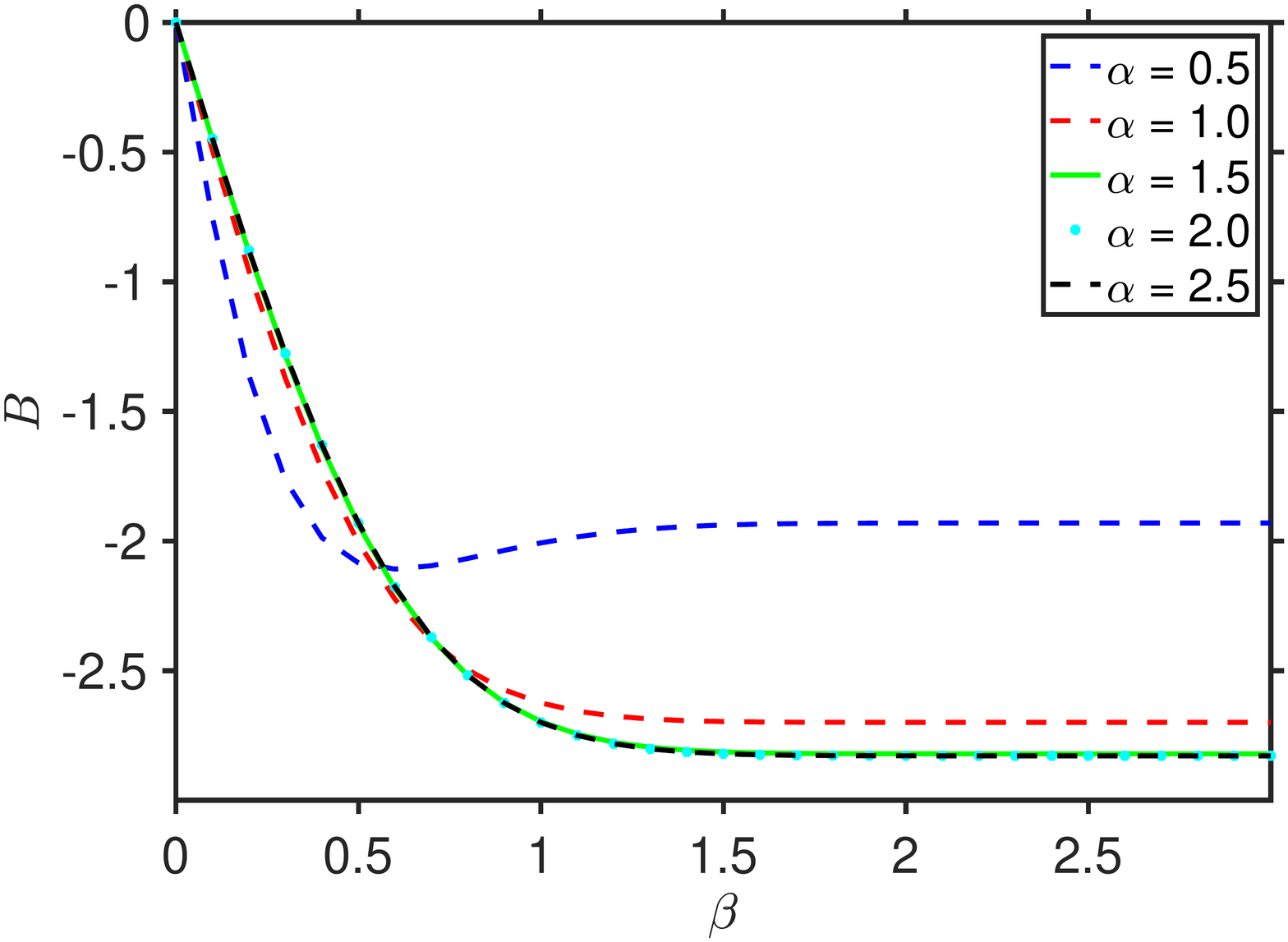}

\caption{The violation of the macroscopic Bell inequality (\ref{eq:chshtime})
for the system depicted in Figure 1, as described in Figure 5. We
show (top) $B$ versus $\beta$ where $\alpha=\beta.$ In the lower
plot, $\beta$ is varied for fixed $\alpha$. A violation is obtained
when $|B|>2$.\textcolor{red}{}\textcolor{red}{}}
\end{figure}

\subsection{Dynamics of the unitary part of the measurement}

In the set-up of Figure 1, two spacelike separated systems display
failure of macroscopic local realism. As the measurements take place,
the systems evolve dynamically over an interval of time. This is a
realistic model because the analyser measurements involve interactions
over a finite time duration. In order to gain insight into how the
\emph{macroscopic} Bell violations arise over the course of the dynamics,
we present plots in terms of the Q function.

The Husimi Q function is defined uniquely as a positive function for
a two-mode quantum state $\rho$ by $Q(\alpha_{0},\beta_{0})=\frac{1}{\pi^{2}}\langle\alpha_{0}|\langle\beta_{0}|\rho|\beta_{0}\rangle|\alpha_{0}\rangle$
where $|\alpha_{0}\rangle$ and $|\beta_{0}\rangle$ are coherent
states for the modes \cite{Husimi-Q-1}. It is a function of four
real variables $x_{A}$, $p_{A}$, $x_{B}$ and $p_{B}$, where $\alpha_{0}=x_{A}+ip_{A}$
and $\beta_{0}=x_{B}+ip_{B}$. The Q function is a quasi-probability
distribution, corresponding to an anti-normal ordering of moments,
and therefore (different to a positive Wigner function) does not provide
a probability distribution for the actual outcomes $X_{A}$, $P_{A},$
$X_{B}$, $P_{B}$ of the measurements \textcolor{black}{$\hat{X}_{A}={\color{red}{\color{blue}{\color{black}\frac{1}{\sqrt{2}}}}}(\hat{a}+\hat{a}^{\dagger}),\hat{X}_{B}={\color{red}{\color{blue}{\color{black}\frac{1}{\sqrt{2}}}}}(\hat{b}+\hat{b}^{\dagger})$},
\textcolor{black}{$\hat{P}_{A}={\color{red}{\color{blue}{\color{black}\frac{1}{i\sqrt{2}}}}}(\hat{a}-\hat{a}^{\dagger}),\hat{P}_{B}={\color{red}{\color{blue}{\color{black}\frac{1}{i\sqrt{2}}}}}(\hat{b}-\hat{b}^{\dagger})$}.
However, the $Q$ function provides at any given time the correct
probability distribution for the outcomes of the \emph{macroscopically
separated} spins $S_{j}^{(B)}$ and $S_{i}^{(A)}$, as we confirm
below.
\begin{figure}[t]
\begin{centering}
\includegraphics[width=1\columnwidth]{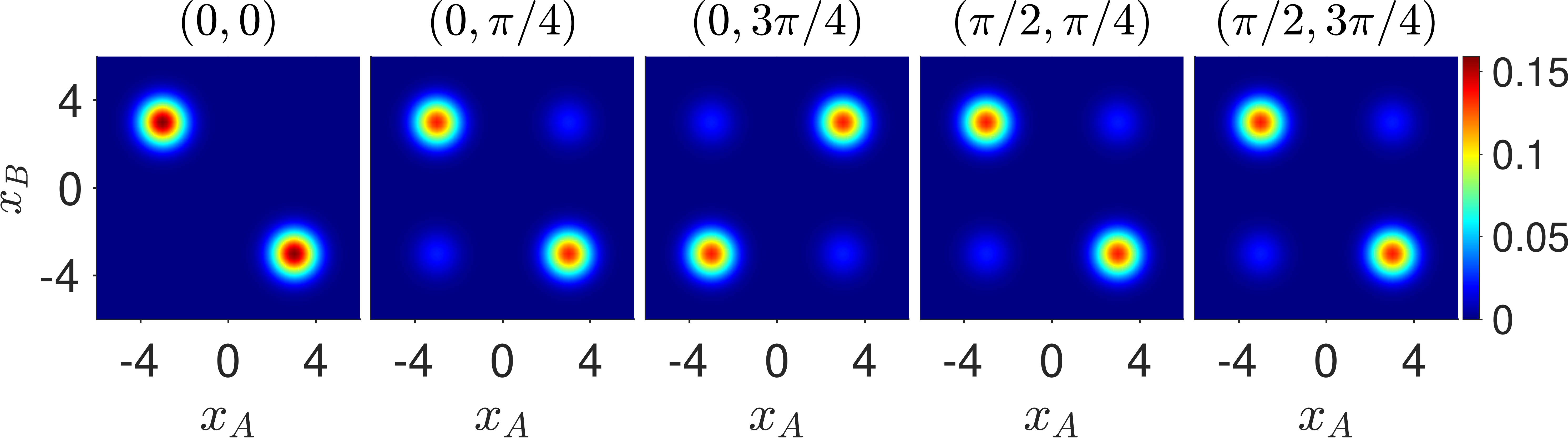}
\par\end{centering}
\smallskip{}

\smallskip{}

\begin{centering}
\includegraphics[width=1\columnwidth]{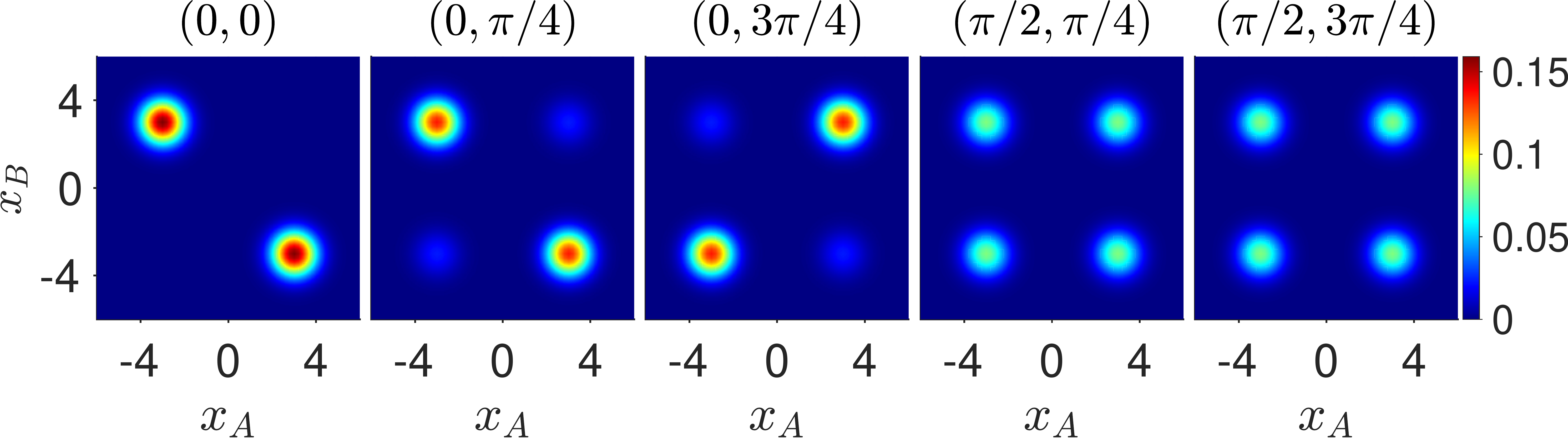}
\par\end{centering}
\begin{centering}
\par\end{centering}
\begin{centering}
\par\end{centering}
\caption{Top: Contour plots of the $Q$ function marginal $Q(x_{A},x_{B})$
for the states created at the times relevant for the violation of
the macroscopic Bell inequality (\ref{eq:chshtime}), as in Figure
5. The system is prepared at $t_{a}=t_{b}=0$ in the entangled Bell
cat state $|\psi_{Bell}\rangle$ (\ref{eq:cat-1}). Lower: The corresponding
plots if the system is prepared in the non-entangled mixture $\rho_{mix}$
(\ref{eq:mixqstate}), which does not violate the Bell inequality.
Here  $\alpha=\beta=3$ and $\Omega=1$.\textcolor{red}{}}
\end{figure}

\begin{figure}[t]
\begin{centering}
\includegraphics[width=1\columnwidth]{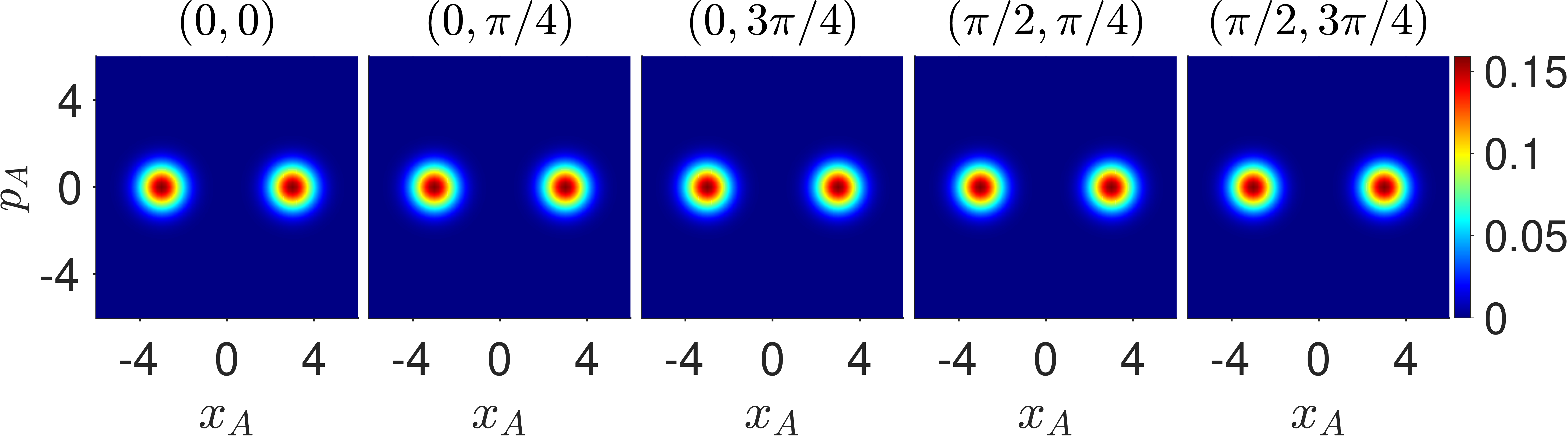}
\par\end{centering}
\begin{centering}
\par\end{centering}
\caption{The two-state nature of the systems created at the specific times
given in Figures 5 and 7 is evident on examining the contour plots
of the marginal $Q(x_{A},p_{A})$. The plots for the Bell cat state
and the mixed state $\rho_{mix}$ are indistinguishable.\textcolor{blue}{}\textcolor{red}{}}
\end{figure}
Assuming the system is prepared in the cat state (\ref{eq:cat-1}),
the state of the system after interaction times $t_{a}$
at site $A$ and $t_{b}$ at site $B$ is given by eq. (\ref{eq:234-1-1-1}).
The Q function at the time $t_{a}$ and $t_{b}$ is $Q(x_{A},x_{B},p_{A},p_{B})=\frac{1}{\pi^{2}}|\langle\beta_{0}|\langle\alpha_{0}|\psi_{f}\rangle|^{2}$.
The marginal for the two $X$ quadratures, given by\textcolor{blue}{}\textcolor{red}{}
\begin{eqnarray}
Q(x_{A},x_{B}) & = & \int dp_{A}dp_{B}Q(x_{A},x_{B},p_{A},p_{B})\label{eq:qmarg}
\end{eqnarray}
\textcolor{red}{}One can also evaluate the marginals for each system
e.g. $Q(x_{A},p_{A}$) by integrating over $x_{B}$ and $p_{B}$.

In Figures 7 and 8, we plot the marginals $Q(x_{A},x_{B})$ and $Q(x_{A},p_{A})$
corresponding to the states created after the interaction times $t_{a}$
and $t_{b}$, as in Figure 5, where there is a violation of the Bell
inequality (\ref{eq:chshtime}). As expected, there is similarity
in the macroscopic limit of the Q function $Q(x_{A},x_{B})$ with
the actual distributions $P(X_{A},X_{B})$, of Figure 5. The additional
noise of order $\hbar$ (scaled to order $\sim1$ in the plots) which
is characteristic of the Q function does not change the probabilities
for the macroscopically distinct outcomes, $S_{A}$ and $S_{B}$.
\begin{figure*}[t]
\begin{centering}
\includegraphics[width=1.8\columnwidth]{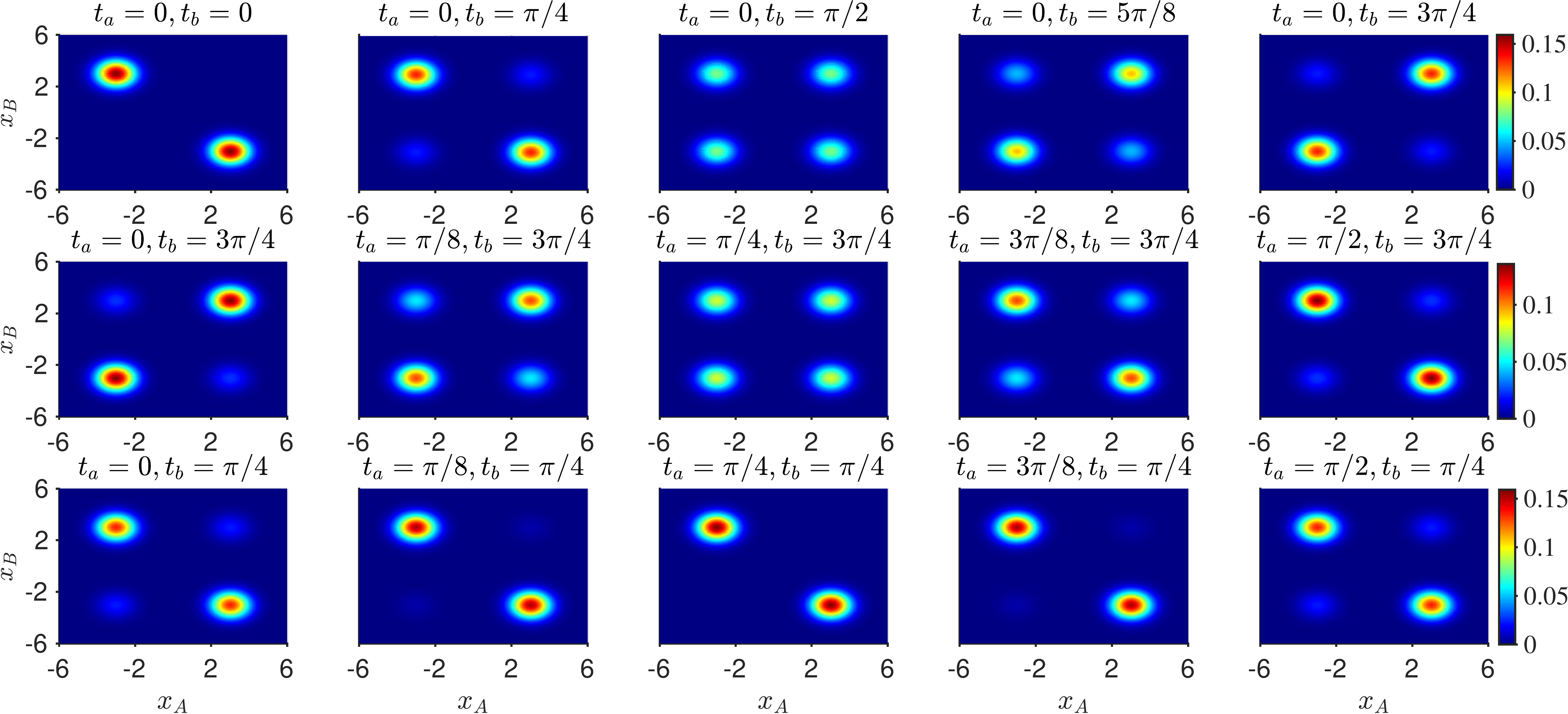}
\par\end{centering}
\begin{centering}
\par\end{centering}
\caption{The entangled cat state $|\psi_{Bell}\rangle$ (\ref{eq:cat-1}) evolves
while being measured at the different sites. We show the contour
plots of $Q(x_{A},x_{B})$ corresponding to the $4$ joint measurements
leading to violation of the Bell inequality (2), as in Figures 5 and
7. \textcolor{red}{}The top sequence shows the dynamics for the
measurements with settings $(t_{a}=0,t_{b}=\pi/4)$ and $(t_{a}=0,t_{b}=3\pi/4)$.
The centre and lower sequences correspond to settings $(t_{a}=\pi/2,t_{b}=3\pi/4)$
and $(t_{a}=\pi/2,t_{b}=\pi/4)$, respectively.\textcolor{red}{{} }Here
$\alpha=\beta=3$ and $\Omega=1$.}
\end{figure*}

\begin{figure*}[t]
\begin{centering}
\includegraphics[width=1.8\columnwidth]{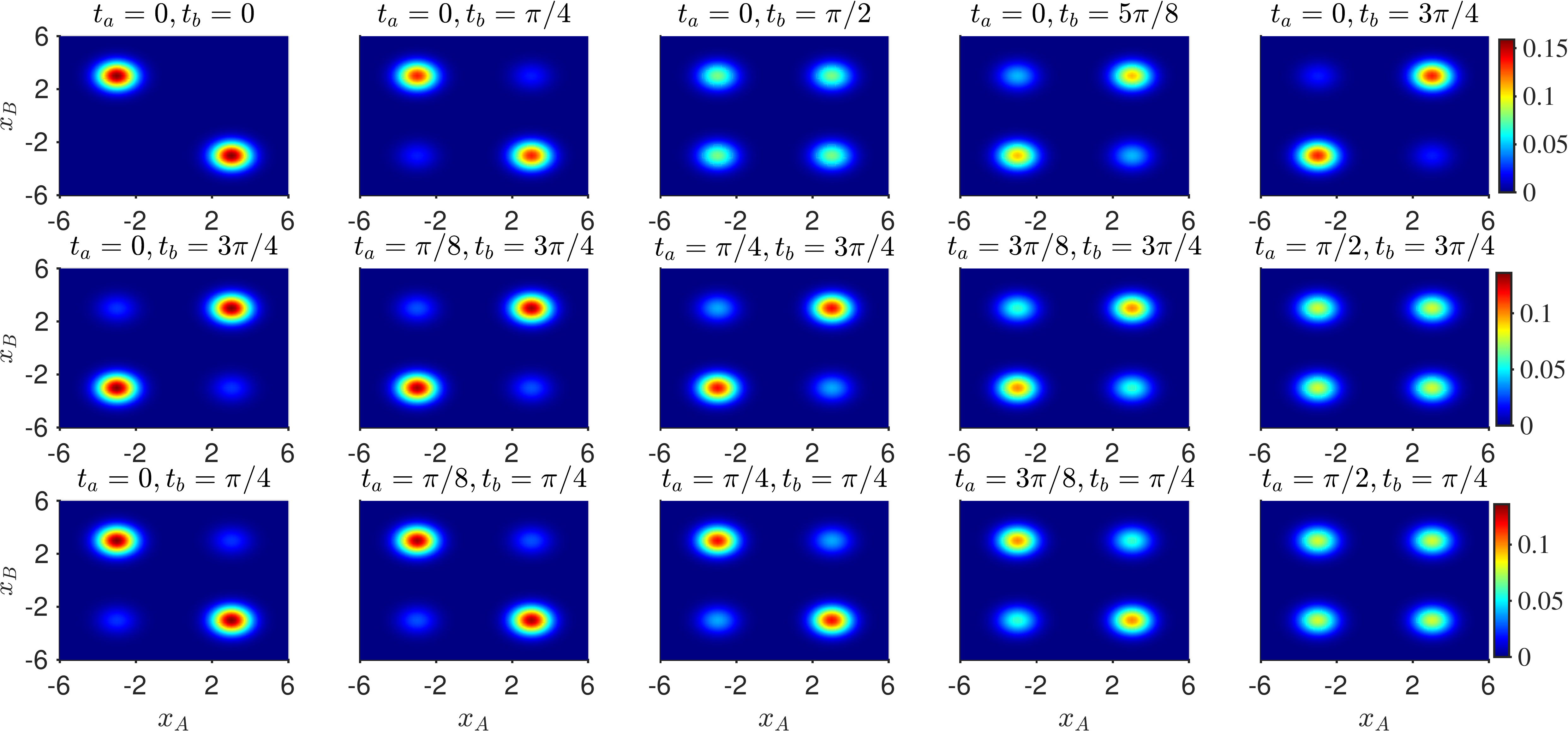}
\par\end{centering}
\begin{centering}
\par\end{centering}
\caption{The plots are as for Figure 9, but here the system evolves from a
nonentangled state $\rho_{mix}$. There is no violation of the Bell
inequality. The plots with $t_{a}=0$ (top) involving a unitary rotation
at only one site are indistinguishable from those of Figure 9. The
centre and lower plots involving transformations at two sites show
a macroscopic difference at time $t_{a}=\pi/(2\Omega)$.\textcolor{red}{}}
\end{figure*}
To investigate the dynamics leading to the violation of macroscopic
local realism (MLR), we give a comparison with the predictions if
the system is initially prepared in the non-entangled mixed state
\begin{eqnarray}
\rho_{mix} & = & \frac{1}{2}(|\alpha\rangle|-\beta\rangle\langle\alpha|\langle-\beta|+|-\alpha\rangle|\beta\rangle\langle-\alpha|\langle\beta|)\nonumber \\
\label{eq:mixqstate}
\end{eqnarray}
The mixture evolves to
\begin{eqnarray}
\rho_{mix}(t_{a},t_{b}) & = & U_{B}(t_{b})U_{A}(t_{a})\rho_{mix}U_{A}^{\dagger}(t_{a})U_{B}^{\dagger}(t_{b})\nonumber \\
\label{eq:mixunit}
\end{eqnarray}
\textcolor{red}{}We plot the Q function of the mixed state $\rho_{mix}$,
eqn. (\ref{eq:mixunit}), in Figure 7 (lower) for the choice of time-settings
that for the state $|\psi_{Bell}\rangle$ lead to a violation of the
Bell inequality (top). The mixed state has the interpretation that
the system is in one \emph{or} other of the quantum states with a
definite outcome for $S_{A}$ and $S_{B}$ at a time $t=0$, as $\alpha$,
$\beta\rightarrow\infty$. Consistent with that interpretation, the
dynamics does not lead to a violation of the Bell inequality. We
see that the Q functions for the initial states ($|\psi_{Bell}\rangle$
and $\rho_{mix}$) at $t=0$ are indistinguishable. In fact, calculation
shows that there is a microscopic difference $\mathcal{C}$, of order
$e^{-|\alpha|^{2}-|\beta|^{2}}$, between these two Q functions.
We also see that the Q functions arising from initial states ($|\psi_{Bell}\rangle$
and $\rho_{mix}$) are indistinguishable if either $t_{a}$ or $t_{b}$
is zero. However, for the time settings where there is evolution (rotation)
at \emph{both} the sites, the final outcomes are macroscopically
different.

The timescales of the dynamics leading to the violation of macroscopic
local realism (MLR) can be visualised if we calculate the full dynamics
associated with the unitary measurements (rotations) (Figures 9
and 10). The predictions for the joint distributions $P(X_{A},X_{B})$
depend only on the absolute value of the times $t_{a}$ and $t_{b}$
of interaction with the measurement apparatus at each site. The same
violation of the Bell inequalities can thus be achieved in different
ways relative to a shared clock. For example, one may measure system
$A$ first, and subsequently measure system $B$; or vice versa.
Alternatively, one may measure by evolving the systems simultaneously.
In Figures 9 and 10, we choose to evolve sequentially, with $B$
first. On comparing the sequences of Q functions for the entangled
state and for the mixture $\rho_{mix}$ in Figures 9 and 10, we observe
the transition from indistinguishable Q functions at $t_{a}=t_{b}=0$
though to a small but noticable difference at times $t_{a}=\pi/8\Omega$
and $t_{b}=\pi/4\Omega$. Finally, at times $t_{a}=\pi/2\Omega$ and
$t_{b}=\pi/4\Omega$, the difference has become macroscopic.

To understand why there is little distinction between the results
for $|\psi_{Bell}\rangle$  and $\rho_{mix}$ if the unitary evolution
acts on the state at one site only ($t_{a}=0$), but not where for
$t_{a}\neq0$ and $t_{b}\neq0$, we calculate the probabilities $P(X_{A},X_{B})$
for each case, in the Appendix. The difference arises from quantum
interference terms damped by $\exp(-x_{A}^{2}-2|\alpha|^{2})$ in
the first case but which are not damped for the entangled state for
certain values of $x_{A}$, in the second case. The effect for large
$\alpha$ and $\beta$ is in direct analogy with the calculations
given for the two-qubit Bell experiment in Section III.A, as expected
since these calculations rely on the orthogonality of the states forming
the qubit.

The plots in Figure 8 show the marginals $Q(x_{A},p_{A})$ and $Q(x_{B},p_{B})$
at each site. These plots highlight the two-state nature of the systems
immediately after the unitary rotations $U_{A}(t_{a})$ and $U_{B}(t_{b})$
for the choice of measurement settings required for the Bell violation.
Identical two-state plots are obtained for the coarsely selected times
given in Figure 9 that show the dynamics during those measurements.
However, as seen from Figure 4, over very much \emph{shorter} time
scales, the unitary dynamics involves a continuous transition, and
the systems cannot be regarded as two-state systems at all times.

\subsection{Conclusions}

It is clear from the summary given in Section II.C that the violations
predicted for the macroscopic Bell inequality (1) and (2) imply failure
of deterministic macroscopic (local) realism (dMR) and macroscopic
local causality (MLC), which we collectively refer to as macroscopic
local realism (MLR). We explain that there is no inconsistency however
with the assumption of weak macroscopic realism (wMR), as introduced
in the Sections I and II.

The peaks of the Q function in the Figures give the values for the
probabilities of the macroscopically distinct outcomes ($+1$ and
$-1$) for the spins, $S_{A}$ and $S_{B}$. The description is of
an entire ensemble, not of an individual system. Therefore the negligible
difference of order $\mathcal{C}\sim e^{-|\alpha|^{2}}$ between the
Q functions of the classical mixture $\rho_{mix}$ (eqn (\ref{eq:mixqstate}))
and the superposition $|\psi_{Bell}\rangle$ (eqn (\ref{eq:cat-1}))
refers to the average over many identically prepared systems, not
to any individual system at any given time.

In what we refer to here as the weak macroscopic realism (wMR) model,
an interpretation is given that goes beyond this. Consider the cat-state
$|\psi_{Bell}\rangle$, which for large $\alpha,$$\beta$ is a superposition
of two states with definite and macroscopically distinct outcomes
for the pointer measurement $\hat{M}^{(A)}$ ($\hat{M}^{(B)}$) of
the spins $S_{A}$ ($S_{B}$). Similar superpositions (\ref{eq:finalstate})
are created after the unitary rotations $U$ at each site, as evident
in Figures 7 and 9. In analogy with the polarising beam splitter or
analyser in a Bell experiment, $U$ brings about a change of basis,
in preparation for a pointer measurement.  The Q function for such
superposition states $|\psi_{pointer}\rangle$ gives macroscopically
distinct peaks corresponding to outcomes $S_{A}$ ($S_{B}$) for $\hat{M}^{(A)}$
($\hat{M}^{(B)}$).

The wMR interpretation is that immediately prior to the measurement
$\hat{M}$ at each site, the local system \emph{was} in one or other
of two states (referred to as $\varphi_{1}$ and $\varphi_{2}$) predetermined
to give one or other of the macroscopically distinguishable outcomes
$+1$ or $-1$ respectively. This assumption is defined in parallel
with that of macroscopic locality of the pointer (MLP): that the measurement
at $B$ does not affect the (macroscopic) spin outcome at $A$ (and
vice versa). In this model, the violation of macrorealism and of
macroscopic local realism arises over the course of the unitary dynamics.
The distinction between predictions for the initial Bell superposition
and a classical mixture of the two pointer eigenstates is quantified
by terms $\mathcal{C}$ of order $\hbar e^{-|\alpha|^{2}}$. The
violations of macro-realism and MLR emerge with the amplification
of the effect of $\mathcal{C}$, over the course of the unitary dynamics
corresponding to the rotation of basis.

In this interpretation, it is \emph{not }assumed that the states $\varphi_{1}$
and $\varphi_{2}$ correspond to $|\alpha\rangle$ and $|-\alpha\rangle$,
nor that $\varphi_{1}$ and $\varphi_{2}$ correspond to quantum states
(indeed, they cannot, as shown in Section V). Nonetheless, the wMR
model asserts the predetermination of the macroscopic prediction.

\section{Leggett-Garg tests of macro-realism and macroscopic local realism}

In this section, we propose tests of macrorealism and of macroscopic
local realism using cat states and the Leggett-Garg inequalities.
For measurements of spin $S_{j}^{(A)}$ made on a single system
$A$ at consecutive times $t_{1}<t_{2}<t_{3}$, macrorealism implies
the Leggett-Garg inequality \cite{weak-solid-state-qubits,jordan_kickedqndlg2-1}\textcolor{red}{}
\begin{equation}
{\color{black}{\color{black}\langle S_{1}^{(A)}S_{2}^{(A)}\rangle+\langle S_{2}^{(A)}S_{3}^{(A)}\rangle-\langle S_{1}^{(A)}S_{3}^{(A)}\rangle}\leq1}\label{eq:lg-ineq}
\end{equation}

\subsection{Leggett-Garg test with a single system}

We first outline the approach using a single system, but as applied
to the cat states. Here, we follow an analysis similar to that given
by Ref. \cite{manushan-cat-lg}. The system $A$ is prepared at time
$t_{1}=0$ in a coherent state $|\alpha\rangle$. The spin $S_{1}$
if measured at this time gives the result $+1$. The result however
is known deterministically, from the preparation. Following preparation,
there is evolution according to the Hamiltonian $H_{NL}^{(A)}$ of
eqn (\ref{eq:ham_NL}), for a time $t_{2}=\pi/4\Omega$. This gives
the state\textcolor{red}{}
\begin{eqnarray}
U_{\pi/8}|\alpha\rangle & = & e^{-i\pi/8}\{(\cos\pi/8|\alpha\rangle+i\sin\pi/8|-\alpha\rangle\}\label{eq:supt2}
\end{eqnarray}
At the time $t_{2}$, after the rotation, we may choose to measure
the value of the spin $S_{2}$. This gives\textcolor{black}{{} $\langle S_{1}S_{2}\rangle=\cos(\pi/4)$}.
\foreignlanguage{australian}{\textcolor{black}{}If \emph{not }measured,
the system continues to evolve according to $H_{NL}$ until the time
$t_{3}=\pi/2\Omega$. The state of the system at $t_{3}$ i}s
\begin{equation}
U_{\pi/4}|\alpha\rangle=\frac{e^{-i\pi/4}}{\sqrt{2}}\{|\alpha\rangle+e^{i\pi/2}|-\alpha\rangle\}\label{eq:supt3}
\end{equation}
\foreignlanguage{australian}{which implies $\langle S_{1}S_{3}\rangle=0$.}\textcolor{red}{}

\selectlanguage{australian}%
To test the Leggett-Garg inequality, one requires to evaluate $\langle S_{2}S_{3}\rangle$.
Here, one assumes macrorealism. We follow the usual approach, and
suppose that an ideal measurement $\hat{M}$ of $S_{2}$ takes place
at time $t_{2}$, which would give a result of either $+1$ or $-1$.
If the result is $\pm1$, then one \emph{assumes} the system was in
the state $|\pm\alpha\rangle$ at the time $t_{2}$, based on the
assumption that the measurement did not disturb the system, and noting
that tomography of the (average) state conditional on the result $\pm1$
would be consistent with the quantum state $|\pm\alpha\rangle.$ For
either state $|\pm\alpha\rangle$, one deduces $\langle S_{2}S_{3}\rangle=\cos(\pi/4)$,
since the system evolves for a further time $\pi/4\Omega$. \foreignlanguage{english}{Thus,
the prediction is }$\langle S_{2}S_{3}\rangle=\cos(\pi/4)$\foreignlanguage{english}{.
Overall, the value for the left-side of the Leggett-Garg inequality
(\ref{eq:lg-ineq}) is $1.4$, which  violates the Leggett-Garg inequality.\textcolor{red}{}}

\selectlanguage{english}%
A criticism of the above approach is that there \emph{is} a change
due to the measurement $\hat{M}$ at time $t_{2}$, and the state
prior to the measurement was different to $|\alpha\rangle$. However,
macrorealism postulates the existence of an ideal measurement $\hat{M}$
where any such effect will have negligible impact on the future dynamics,
as the system becomes macroscopic ($\alpha\rightarrow\infty$). This
would not change the prediction of a violation, if macrorealism is
correct.

The dynamics as depicted in terms of the Q function is given for the
sequence of times $t_{1}<t_{2}<t_{3}$ in Figure 11. There are two
cases to compare: whether a measurement $\hat{M}$ is performed at
time $t_{2}$, or not. Details are given in the Appendix. The lower
sequence plots where a measurement is performed at time $t_{2}$,
assuming that for the ideal measurement this is done in such a way
to instigate a collapse into one or other of the coherent states,
as above. The Q functions (\ref{eq:q2-2}) and (\ref{eq:q2}) describe
the states for an ensemble of systems $A$ immediately after the time
$t_{2}$, if a measurement has or has not taken place at time $t_{2},$
respectively. The two Q functions differ by the term
\begin{equation}
\mathcal{C}=-\frac{e^{-p_{A}^{2}-x_{A}^{2}}}{\sqrt{2}\pi}e^{-\alpha^{2}}\sin(2p_{A}\alpha_{0})\label{eq:c-term}
\end{equation}
This term is negligible for even moderate $\alpha\sim1.5$ and there
is no distinguishable difference between the functions at the time
$t_{2}$ (refer Figure 11).  Despite the negligible difference between
the (average) states at time $t_{2}$, the states at time $t_{3}$
are macroscopically different. The Q function model is consistent
with the Leggett-Garg premise that a measurement exists which has
increasingly negligible effect on the immediate state of the system
as the system becomes larger \cite{legggarg-1}. This is because $\mathcal{C}$
is proportional to $e^{-\alpha^{2}}$. In this model, the Leggett-Garg
inequalities are violated because the small effect nonetheless impacts
the future dynamics at a macroscopic level. 

\begin{figure}[t]
\begin{centering}
\includegraphics[width=0.8\columnwidth]{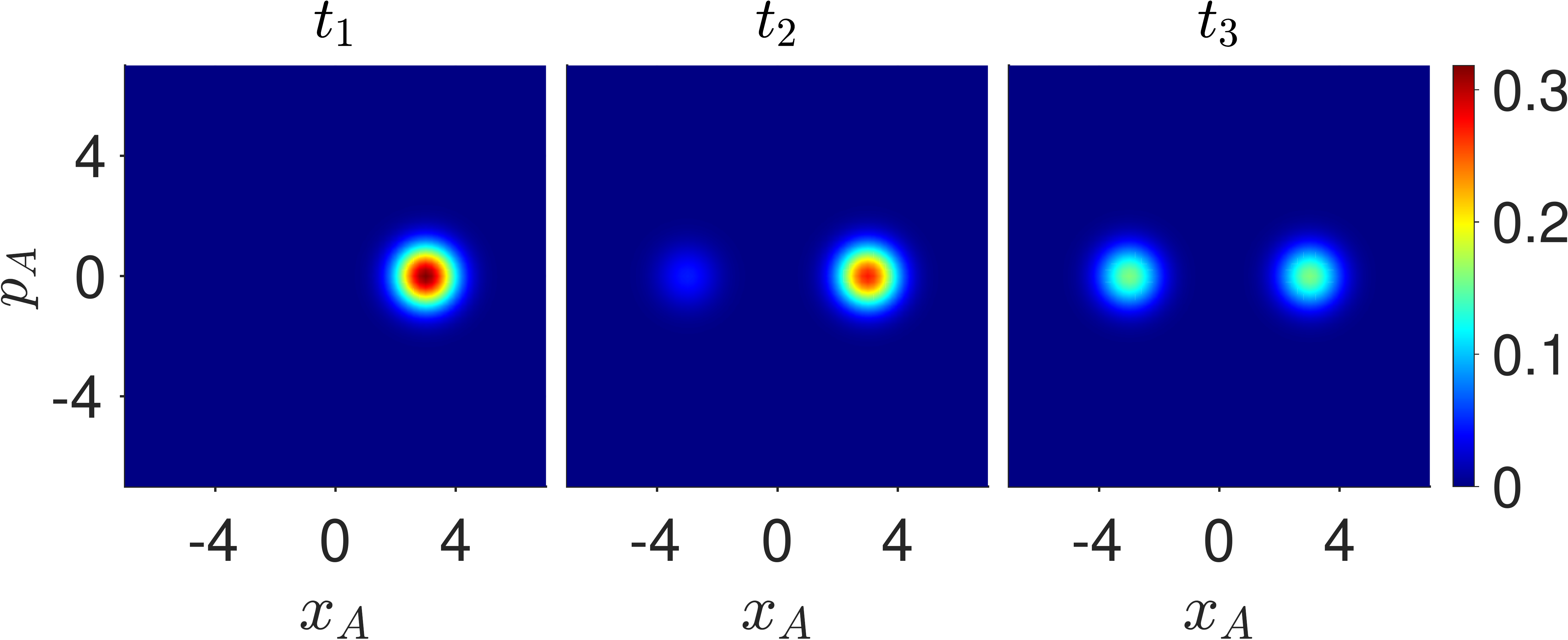}
\par\end{centering}
\smallskip{}
\smallskip{}

\begin{centering}
\includegraphics[width=0.8\columnwidth]{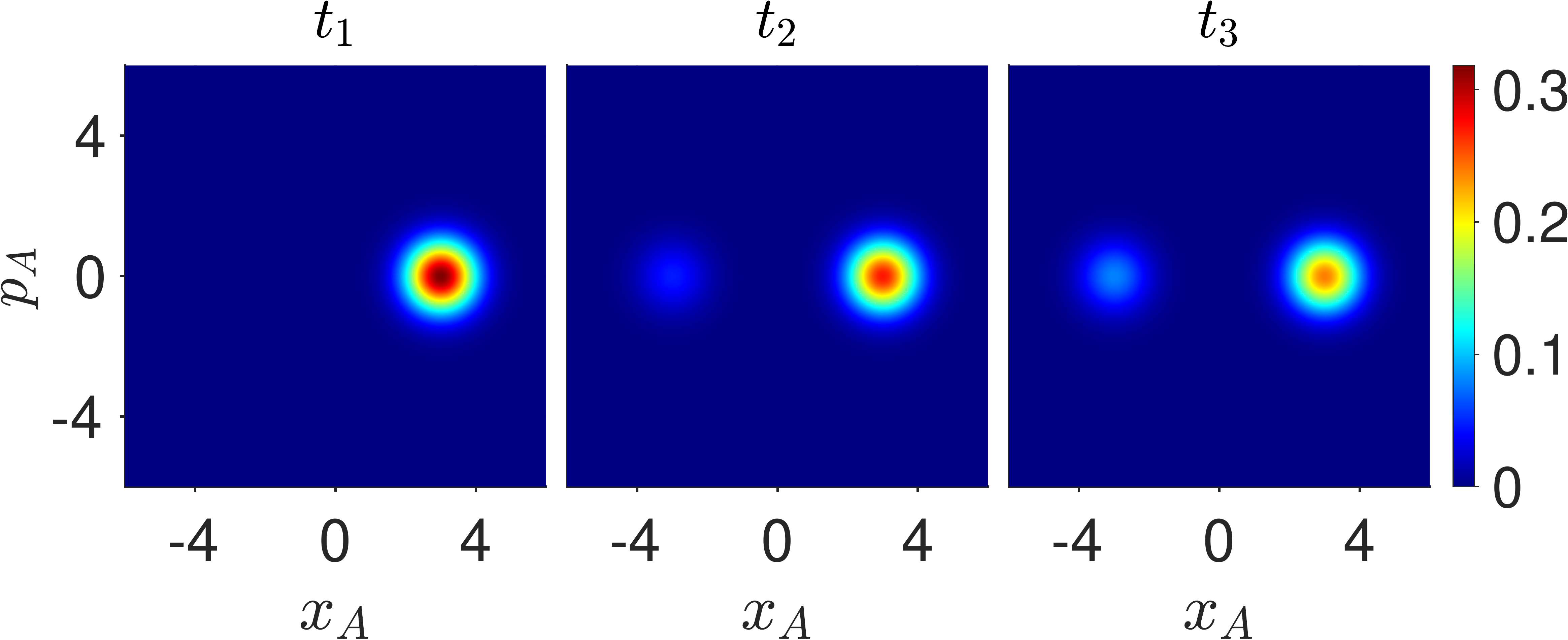}
\par\end{centering}
\begin{centering}
\par\end{centering}
\begin{centering}
\par\end{centering}
\caption{The Q function $Q(x_{A},p_{A})$ for times $t_{1}$, $t_{2}$ and
$t_{3}$ ($\alpha=3$). The top sequence assumes no measurement is
made at $t_{2}$.\textcolor{red}{{} }The lower sequence assumes a measurement
is made at $t_{2}$, of the type that collapses the system into one
or other of two coherent states. While the difference in the Q function
at time $t_{2}$ due to measurement is negligible, this leads to distinct
dynamics, and a macroscopic difference at time $t_{3}$.\textcolor{red}{}\textcolor{red}{}\textcolor{blue}{}\textcolor{red}{}}
\end{figure}

A challenge for testing macrorealism is to overcome the criticism
that the ideal measurement cannot be realised. While arguments can
be made that the system at $t_{2}$ was in a state sufficiently close
to $|\alpha\rangle$ or $|-\alpha\rangle$, if macrorealism is correct
\cite{NSTmunro-1}, it becomes difficult to exclude that the system,
in a theory alternative to quantum mechanics, is described by some
other state at time $t_{2}$, prior to measurement. This motivates
the next section.

\subsection{Bipartite Leggett-Garg tests}

\subsubsection{Violation of inequalities: idealised measurement at $B$}

We next analyse a different model for the ideal measurement that occurs
at time $t_{2}$, where strong justification can be given that the
measurement at $t_{2}$ is nondisturbing. We will see that the predictions
of a violation of the Leggett-Garg inequality (\ref{eq:lg-ineq})
are unchanged. Consider the set-up of Figure 2 where the system is
prepared in a Bell cat state
\begin{equation}
|\psi_{Bell}\rangle=\mathcal{N}(|\alpha\rangle|-\beta\rangle-|-\alpha\rangle|\beta\rangle)\label{eq:cat-1-3}
\end{equation}
as in eqn. (\ref{eq:cat-1}). We propose a test of the Leggett-Garg
inequality (\ref{eq:lg-ineq}) for system $A$, where the second system
$B$ is used to perform measurements on system $A$. After preparation,
both systems $A$ and $B$ interact with the nonlinear medium as given
by (\ref{eq:state5}) at the respective locations and evolve according
to the same shared clock. At the later time $t_{2}=\pi/4\Omega$,
the overall state is
\begin{eqnarray}
U_{\pi/8}^{(A)}U_{\pi/8}^{(B)}|\psi_{Bell}\rangle & = & \mathcal{N}e^{-i\pi/4}(|\alpha\rangle|-\beta\rangle-|-\alpha\rangle|\beta\rangle)\nonumber \\
\label{eq:bellunitary}
\end{eqnarray}
\textcolor{red}{}The system remains in a Bell cat state and there
is a perfect anticorrelation between the spin results $S_{2}^{(A)}$
and $S_{2}^{(B)}$ at $A$ and $B$. \textcolor{red}{}Hence, a
measurement of $S_{i}^{(A)}$ can be made noninvasively, by performing
a measurement of the spin $S_{i}^{(B)}$, where $i=1,2$. The assumption
of noninvasiveness is justified by that of\emph{ }locality: It is
assumed that the outcome of the measurement on the spacelike separated
system $A$ is independent of the choice of duration of the unitary
evolution at $B$.

One requires to measure $\langle S_{1}^{(A)}S_{2}^{(A)}\rangle$.
One infers the result for $S_{1}^{(A)}$ by measuring $S_{1}^{(B)}$.
A measurement is then made of $S_{2}^{(A)}$ at the time $t_{2}$,
assuming such a measurement will accurately reflect the macroscopic
value of the spin $S_{2}^{(A)}$of system $A$ immediately prior to
the time $t_{2}$. Thus \textcolor{red}{}\textcolor{black}{$\langle S_{1}^{(A)}S_{2}^{(A)}\rangle=-\langle S_{1}^{(B)}S_{2}^{(A)}\rangle$}.
To obtain the prediction for \textcolor{black}{$\langle S_{1}^{(A)}S_{2}^{(A)}\rangle$},
one argues as follows. The result of the first measurement at time
$t_{1}$ at $A$ (as measured at $B$) is either $+1$ or $-1$. Assuming
the measurement at $B$ is such to cause a ``collapse'' so that
the system at $A$ is then either in the coherent state $|\alpha\rangle$
or $|-\alpha\rangle$, the system evolves as given by eq. (\ref{eq:supt2})
and $\langle S_{1}^{(A)}S_{2}^{(A)}\rangle=\cos(\pi/4)$. \textcolor{black}{Similarly,
$\langle S_{1}^{(A)}S_{3}^{(A)}\rangle=-\langle S_{1}^{(B)}S_{3}^{(A)}\rangle=\cos(\pi/2)=0$.}

One also requires to measure $\langle S_{2}^{(A)}S_{3}^{(A)}\rangle$.
The result for $S_{2}^{(A)}$ at time $t_{2}$ is inferred by measuring
$S_{2}^{(B)}$ at time $t_{2}$. However, to measure $S_{2}^{(A)}$
this way, we use state (\ref{eq:bellunitary}), which means that \emph{no}
measurement is made on system $B$ (or $A$) at the earlier time $t_{1}$.
The measurement at time $t_{2}$ implies either $+1$ or $-1$, for
the spin of $A$ at time $t_{2}$, and hence we obtai\textcolor{black}{n
the prediction for $\langle S_{2}^{(A)}S_{3}^{(A)}\rangle=\cos(\pi/4)$.
}This gives the left side of the Leggett-Garg inequality (\ref{eq:lg-ineq})
as $1.4$ and a violation is predicted, as in Section IV.A.\textcolor{red}{}\textcolor{blue}{}\textcolor{blue}{}\textcolor{red}{}

The analysis also implies violation of the Leggett-Garg-Bell inequality
(\ref{eq:lg-double-three-1}) presented in Section II. We have taken
the anti-correlated states (\ref{eq:cat-1}) (and (\ref{eq:cat-1-3})),
so that the Leggett-Garg inequality (\ref{eq:lg-ineq}) is written
\begin{equation}
{\color{black}{\color{black}-\langle S_{1}^{(B)}S_{2}^{(A)}\rangle-\langle S_{2}^{(B)}S_{3}^{(A)}\rangle+\langle S_{1}^{(B)}S_{3}^{(A)}\rangle}\leq1}.\label{eq:bipartite-neg}
\end{equation}
It is convenient to consider the correlated Bell state (\ref{eq:cat-1})
(and (\ref{eq:cat-1-3})) given by $\beta\rightarrow-\beta$ so that
the values of the spins $S_{i}^{(A)}$ and $S_{i}^{(B)}$ are equal.
The Leggett-Garg inequality (\ref{eq:lg-ineq}) becomes
\begin{equation}
{\color{black}{\color{black}\langle S_{1}^{(B)}S_{2}^{(A)}\rangle+\langle S_{2}^{(B)}S_{3}^{(A)}\rangle-\langle S_{1}^{(B)}S_{3}^{(A)}\rangle}\leq1}\label{eq:lg-ineq-1-1}
\end{equation}
which may also be measured as
\begin{equation}
{\color{black}{\color{black}\langle S_{1}^{(A)}S_{2}^{(B)}\rangle+\langle S_{2}^{(B)}S_{3}^{(A)}\rangle-\langle S_{1}^{(B)}S_{3}^{(A)}\rangle}\leq1}\label{eq:lg-ineq-1-1-1}
\end{equation}
This reduces to the Leggett-Garg-Bell inequality (\ref{eq:lg-double-three-1})
of Section I. Thus, $E(t_{1},t_{2})=\langle S_{1}^{(A)}S_{2}^{(B)}\rangle$,
$E(t_{2},t_{3})=\langle S_{2}^{(B)}S_{3}^{(A)}\rangle$ and $E(t_{1},t_{3})=\langle S_{1}^{(A)}S_{3}^{(A)}\rangle$.

\subsubsection{Predictions for the realistic measurement}

The above calculation assumes that the measurement at $B$ ``collapses''
the system at $A$ into one or other of the coherent states, $|\alpha\rangle$
or $|-\alpha\rangle$, at the time of the measurement, $t_{2}$. The
calculations based on that assumption can now be rigorously validated,
since we give a specific proposal, that the measurement at site $B$
be performed as a quadrature phase amplitude measurement $\hat{X}_{B}$.
The spin $S_{j}^{(B)}$is the sign of the outcome $X_{B}$. In this
section, we carry out a complete analysis by evaluating $P(X_{A},X_{B})$
as in Section III.C. The results shown in Figures 12 and 13 confirm
the prediction of the violation of the Leggett-Garg-Bell inequality
(\ref{eq:lg-ineq}) and (\ref{eq:lg-double-three-1}) for large $\alpha,$
$\beta$.
\begin{figure}[t]
\begin{centering}
\includegraphics[width=1\columnwidth]{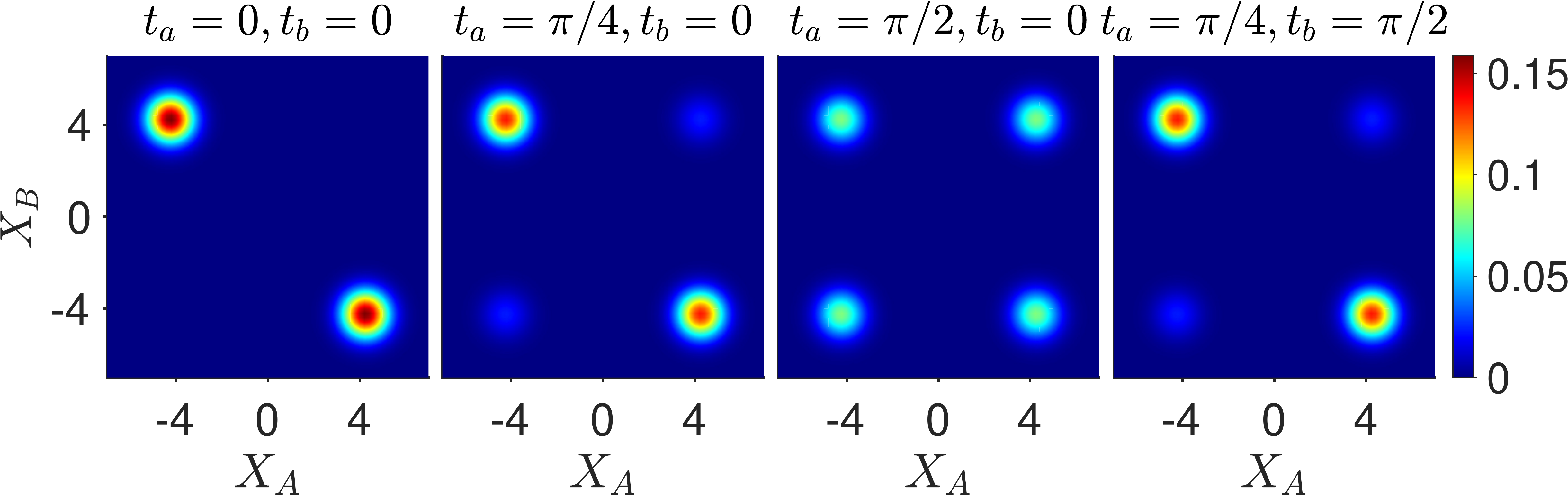}
\par\end{centering}
\begin{centering}
\par\end{centering}
\caption{Distributions giving the violation of the bipartite Leggett-Garg-Bell
inequality (\ref{eq:bipartite-neg}). Two systems $A$ and $B$ are
prepared in the entangled cat state $|\psi_{Bell}\rangle$ (\ref{eq:cat-1})
and then evolve according to the unitary transformations $U_{A}(t_{a})$
and $U_{B}(t_{b})$ (\ref{eq:state5}) as in Figure 2. We show the
contour plots of the joint probability of $P(X_{A},X_{B})$ for the
measurements $\hat{X}_{A}$ and $\hat{X}_{B}$ made at the times $t_{a}$
and $t_{b}$, with $\alpha=\beta=3$. The last three distributions
give the moments $\langle S_{1}^{(B)}S_{2}^{(A)}\rangle$, $\langle S_{1}^{(B)}S_{3}^{(A)}\rangle$
and $\langle S_{2}^{(B)}S_{3}^{(A)}\rangle$. Here, $t_{1}=0$, $t_{2}=\pi/4\Omega$,
and $t_{3}=\pi/2\Omega$. \textcolor{red}{}Similar distributions
and violations are obtained for the bipartite Leggett-Garg-Bell inequality
(\ref{eq:lg-double-three-1}) for the correlated Bell state, where
$X_{B}\rightarrow-X_{B}$.\textcolor{red}{}}
\end{figure}
\begin{figure}[t]
\includegraphics[width=0.7\columnwidth]{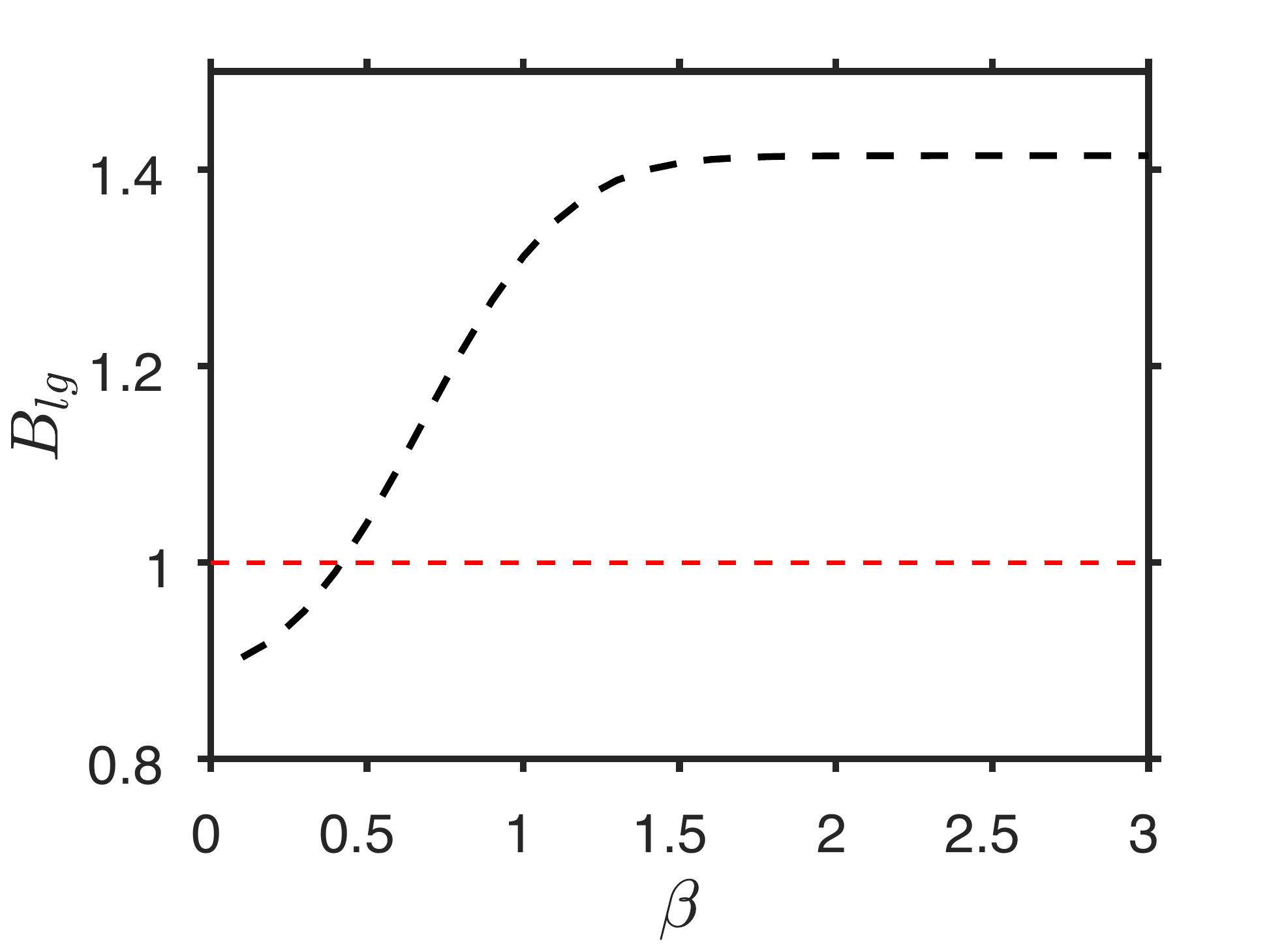}

\includegraphics[width=0.75\columnwidth]{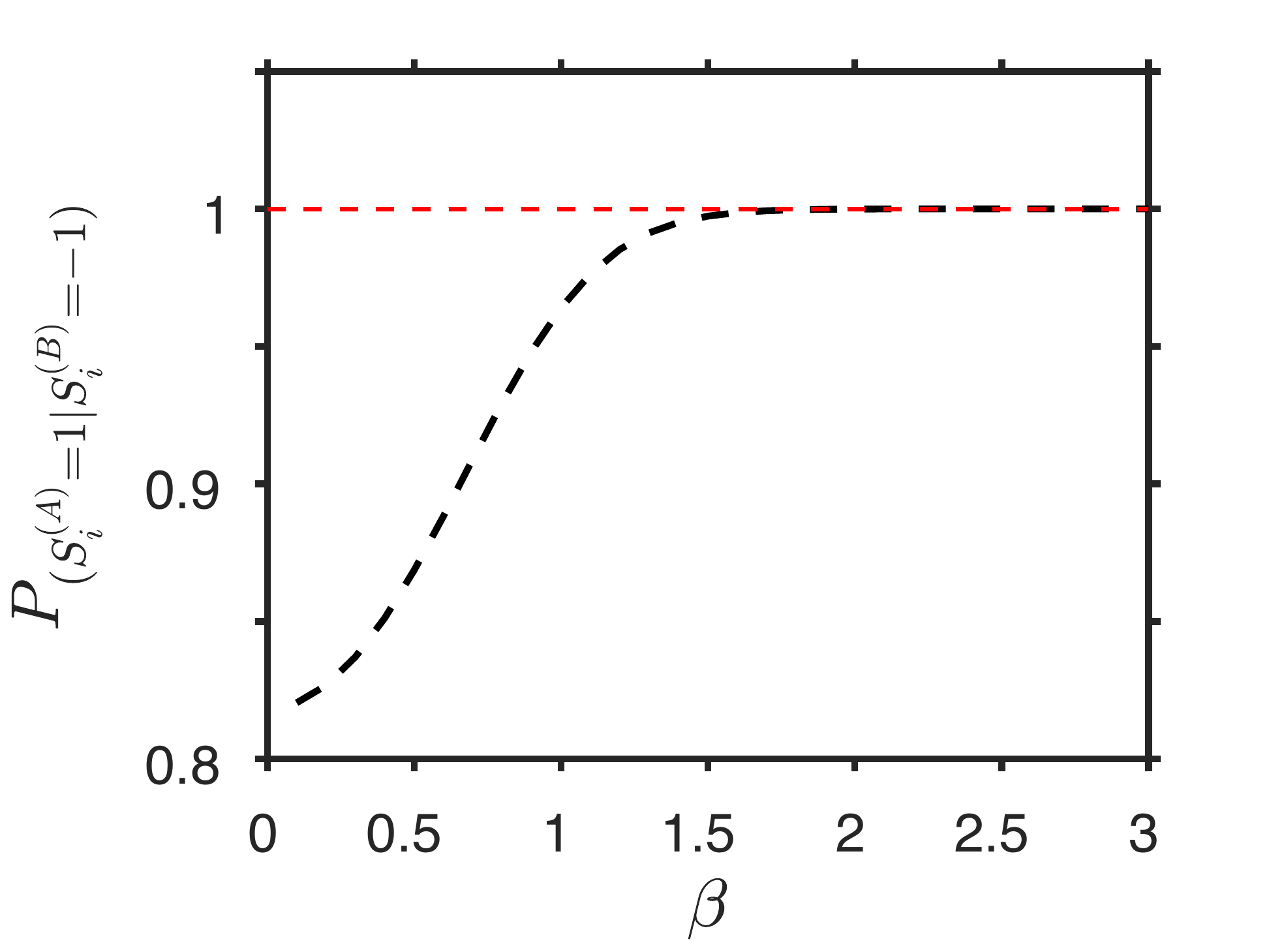}

\caption{The violation of the Leggett-Garg-Bell inequality (\ref{eq:bipartite-neg})
and (\ref{eq:lg-double-three-1}) for the system depicted in Figures
2 and 12. We plot $B_{lg}$ given by (\ref{eq:B-lg}) versus $\beta$
for the state $|\psi_{Bell}\rangle$ (\ref{eq:cat-1}), where $\alpha=\beta.$
Violation is obtained when $B_{lg}>1$. The verification of $\langle S_{i}^{(A)}S_{j}^{(A)}\rangle=-\langle S_{i}^{(B)}S_{j}^{(A)}\rangle$
for $i=1,2$ is given by examination of the conditional distribution
$P(S_{i}^{(A)}=1|S_{i}^{(B)}=-1)$ as shown. \textcolor{red}{}\textcolor{red}{}The
same results are obtained for the Bell state (\ref{eq:cat-1}) with
$\alpha=-\beta$, on defining $B_{lg}=\langle S_{1}^{(A)}S_{2}^{(B)}\rangle-\langle S_{1}^{(B)}S_{3}^{(A)}\rangle+\langle S_{2}^{(B)}S_{3}^{(A)}\rangle$
and considering $P(S_{i}^{(A)}=1|S_{i}^{(B)}=1)$.\textcolor{red}{}}
\end{figure}

The dynamics of the measurement and its disturbance is visualised,
by plotting sequences of the $Q$ function (Figure 14). In order to
measure $\langle S_{1}^{(A)}S_{2}^{(A)}\rangle$ or $\langle S_{1}^{(A)}S_{3}^{(A)}\rangle$,
via the measurable moments $\langle S_{1}^{(B)}S_{2}^{(A)}\rangle$
and $\langle S_{1}^{(B)}S_{3}^{(A)}\rangle$, one measures the spin
$B$ at $t_{1}=0$ (top and centre sequences). In order to measure
$\langle S_{2}^{(A)}S_{3}^{(A)}\rangle$ via the measurable moment
$\langle S_{2}^{(B)}S_{3}^{(A)}\rangle$, one measures the spin $B$
at time $t_{2}$ (with no measurement at $t_{1}$). The measurements
are made by \emph{stopping} the evolution at $B$ at the time $t_{1}$,
or $t_{2}$, respectively.
\begin{figure*}[t]
\begin{centering}
\includegraphics[width=1.8\columnwidth]{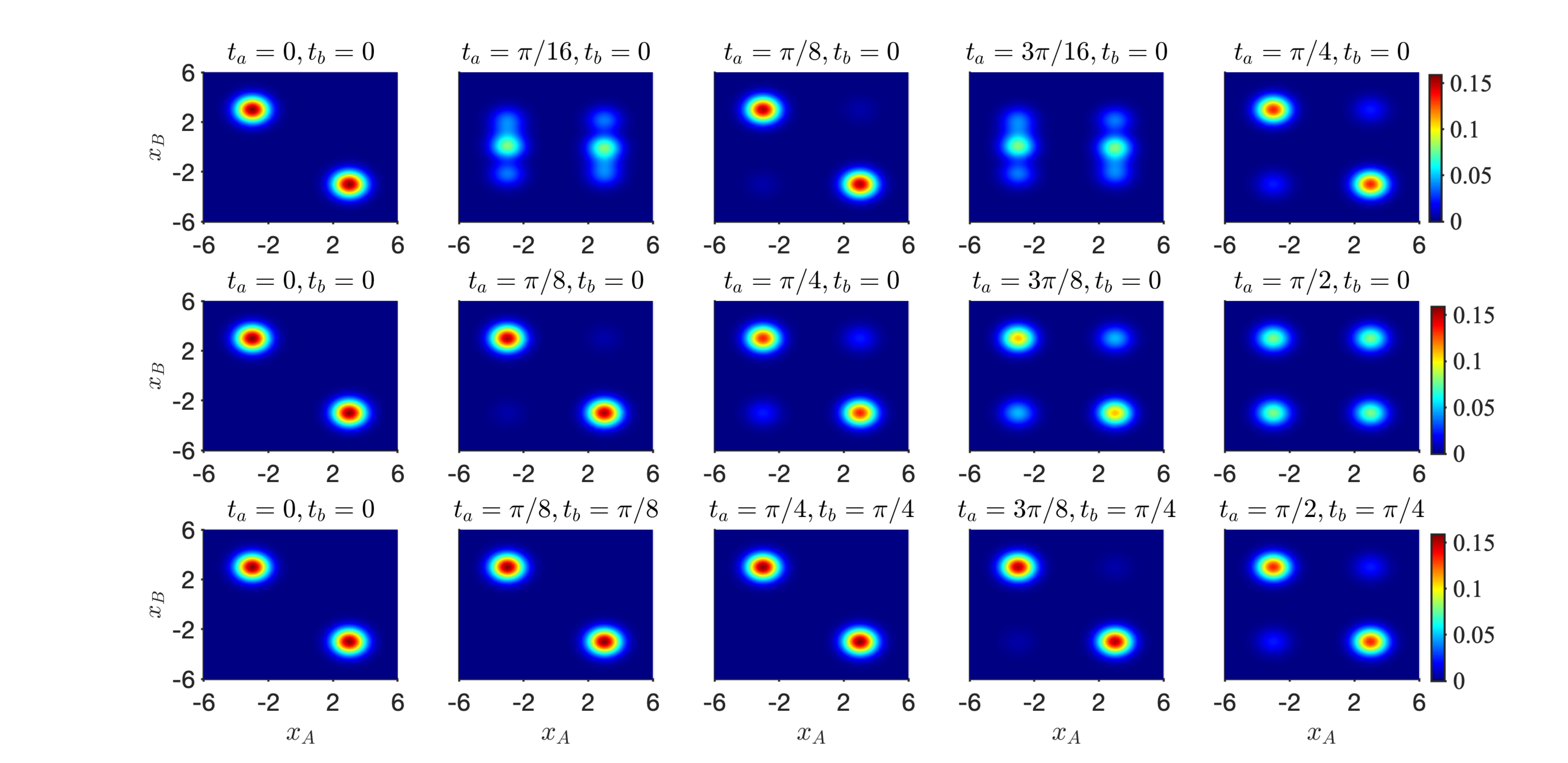}
\par\end{centering}
\caption{Q function dynamics as the entangled cat state $|\psi_{Bell}\rangle$
evolves through the three measurement sequences of the bipartite Leggett-Garg
test. Plots show the marginal distribution $Q(x_{A},x_{B})$ after
evolving for times $t_{a}$ and $t_{b}$ at the sites $A$ and $B$.
The subsystems evolve acording to the same shared clock, with $t_{a}=t_{b}$,
up to the time of measurement at the site $B$. Here, $t_{1}=0$,
$t_{2}=\pi/4\Omega$, and $t_{3}=\pi/2\Omega$. The top, centre and
lower rows give a sequence of snapshots corrresponding to the measurement
of $\langle S_{1}^{(B)}S_{2}^{(A)}\rangle$, $\langle S_{1}^{(B)}S_{3}^{(A)}\rangle$
and $\langle S_{2}^{(B)}S_{3}^{(A)}\rangle$ respectively. The results
agree with those of Figure 11.\textcolor{red}{}\textcolor{red}{}}
\end{figure*}

The top two sequences of Figure 14 show measurement of $\langle S_{1}^{(A)}S_{2}^{(A)}\rangle$
or $\langle S_{1}^{(A)}S_{3}^{(A)}\rangle$, where the system at $B$
is frozen at $t_{1}=t_{b}=0$, so that information about the state
of the system $A$ at time $t_{1}$ is stored at that site $B$, and
$A$ continues to evolve. We see from the plot of $Q(x_{A},x_{B})$
for $t_{b}=0$ and $t_{a}=t_{3}$ (last plot of the centre sequence)
that $\langle S_{1}^{(B)}S_{3}^{(A)}\rangle=0$. Similarly, the
plot for $t_{b}=0$ and $t_{a}=t_{2}$ (last plot of the top sequence)
shows $\langle S_{1}^{(B)}S_{2}^{(A)}\rangle=-0.707$. The lower
sequence of Figure 14 corresponds to a measurement at site $B$ made
at time $t_{2}$, in order to evaluate $\langle S_{2}^{(B)}S_{3}^{(A)}\rangle=-\langle S_{2}^{(A)}S_{3}^{(A)}\rangle$.
The evolution at $B$ ceases at $t_{2}$ so that the value for $S_{2}^{(A)}$
can be measured. The system $A$ continues to evolve, until $t_{3}$.
The correlations giving $\langle S_{2}^{(B)}S_{3}^{(A)}\rangle=-\langle S_{2}^{(A)}S_{3}^{(A)}\rangle$
are indicated by the final plot of the lower sequence. We find $\langle S_{2}^{(B)}S_{3}^{(A)}\rangle=-0.707$
and hence $\langle S_{2}^{(A)}S_{3}^{(A)}\rangle=0.707$. The exact
values for the correlations are evaluated numerically by integration
of the joint distributions $P(X_{A},X_{B})$ at the specified times.
The calculation given in the Appendix leads to\textcolor{red}{}
\begin{eqnarray}
{\color{red}{\color{black}B{\color{red}}_{lg}}} & = & -\{\langle S_{1}^{(B)}S_{2}^{(A)}\rangle-\langle S_{1}^{(B)}S_{3}^{(A)}\rangle+\langle S_{2}^{(B)}S_{3}^{(A)}\rangle\}\nonumber \\
 & = & \frac{\sqrt{2}erf(\sqrt{2}\alpha)erf(\sqrt{2}\beta)}{{\color{blue}{\color{black}\{1-e^{-2\left|\alpha\right|^{2}-2\left|\beta\right|^{2}}\}}}}\label{eq:B-lg}
\end{eqnarray}
where $erf$ is the error function. Results are shown in Figure 13,
in agreement with the simple approach of the last section.

The pictures of Figure 11 are also justified for this measurement
procedure. The impact of whether the measurement takes place at time
$t_{2}$, or not, is seen to be minimal if we examine the marginal
(reduced) state for the system A, immediately after time $t_{2}$.
This is consistent with the idealised model of the measurement given
in Figure 11. If there was no measurement at $t_{2}$, and the measurement
was made at time $t_{1}$, then the system $A$ at time $t_{2}$ is
in a superposition of the two coherent states. On the other hand,
if a measurement is made at time $t_{2}$, then the marginal (reduced)
state for $A$ is the mixture, since no preparation took place at
$t_{1}$ and the statistics for the system is given by (\ref{eq:bellunitary}).
The marginal Q functions in each case are (\ref{eq:q2}) and (\ref{eq:q2-2}),
given as the second plots of the two sequences in Figure 11, which
give indistinguishable predictions for large $\alpha$, justifying
the arguments that this is an (apparently) non-disturbing measurement.

Despite the negligible difference in the marginal Q functions for
system $A$ at the time $t_{2}$, a macroscopic difference in the
\emph{joint correlations} measured between $A$ and $B$ emerges at
the later time $t_{3}$ (and indeed, has already emerged at the time
$t_{2}$). The correlations shown in Figures 12 and 14 imply results
for the two-time moments corresponding exactly to those shown in the
final plots of the two sequences of Figure 11. The macroscopic difference
between the final plots of the two sequences in Figure 11 \emph{arises
over the timescale of the unitary evolution that determines the measurement
}$-$ this is the evolution seen in the first three plots of the centre
and lower sequence, of Figure 14. The subsequent dynamics at $A$
after $t_{2}$ leads to the macroscopic difference in correlations
seen in the final plots of the sequences of Figure 11. On very \emph{short}
timescales (evident at $t_{a}=\pi/16\Omega$ of Figure 14) as the
unitary rotation takes place, we note the system cannot be considered
a two-state system.\textcolor{blue}{}

\section{Delayed collapse}

\subsection{Delaying the collapse stage of the measurement}

We now clarify a possible point of confusion about the timing of the
final readout (the ``collapse'' stage) of the measurement at $B$,
for the Leggett-Garg-Bell proposal of Section IV.B. The measurement
of $S_{1}^{(A)}$ is done by measuring $S_{1}^{(B)}$ at time $t_{1}=0$
and inferring from the correlation between the two systems, $A$ and
$B$. The measurement of $S_{2}^{(B)}$ comes in two stages: First,
it is necessary to \emph{stop} the unitary interaction of system $B$
at the time $t_{1}=0$, so that information from the correlation can
be stored. The second stage of the measurement at $B$ constitutes
the irreversible ``collapse'', where there is a coupling to a detector.
While the timing of the first stage is crucial, the timing of the
``collapse'' stage at $B$ is immaterial.

To clarify, we compare where the collapse at $B$ has, or has not,
occurred prior to the time of the (collapse) measurement of $X_{A}$
at $A$. First, let us assume the collapse at $B$ has \emph{not }occurred.
Here, the two systems $A$ and $B$ are prepared in the entangled
Bell state at time $t_{1}=0$. System $B$ does not evolve further
($t_{b}=0$), while system $A$ evolves for a time $t_{a}>0$. The
state formed after the evolution at $A$ is the superposition
\begin{equation}
U_{A}(t_{a})|\psi_{Bell}\rangle=\mathcal{N}U_{A}(t_{a})\{|\alpha\rangle_{a}|-\beta\rangle_{b}-|-\alpha\rangle_{a}|\beta\rangle_{b}\}\label{eq:cat-1-3-1}
\end{equation}
The final joint distribution $P(X_{A},X_{B})_{sup}$ that describes
the statistics at the later time $t_{a}$, assuming there has been
no prior collapse at $B$, is calculated straightforwardly for this
superposition state, as shown in the Appendix.

\begin{figure*}[t]
\begin{centering}
\includegraphics[width=1.6\columnwidth]{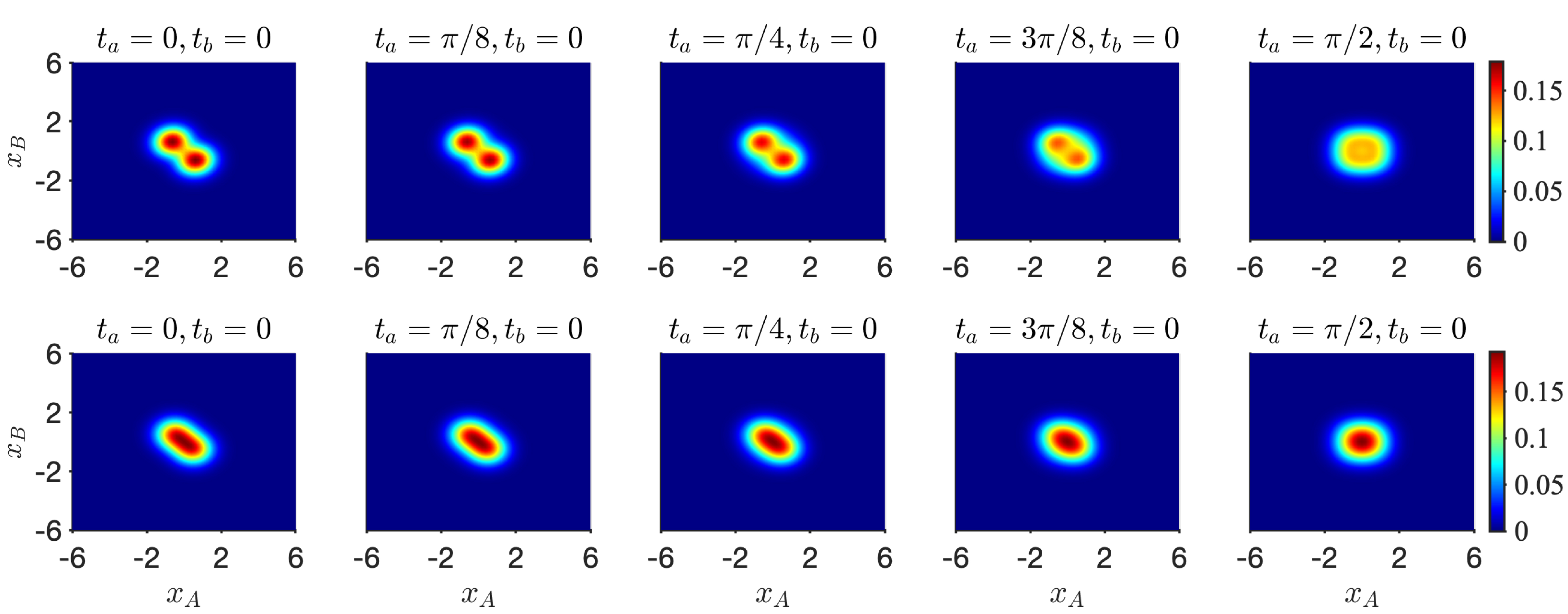}
\par\end{centering}
\begin{centering}
\includegraphics[width=1.6\columnwidth]{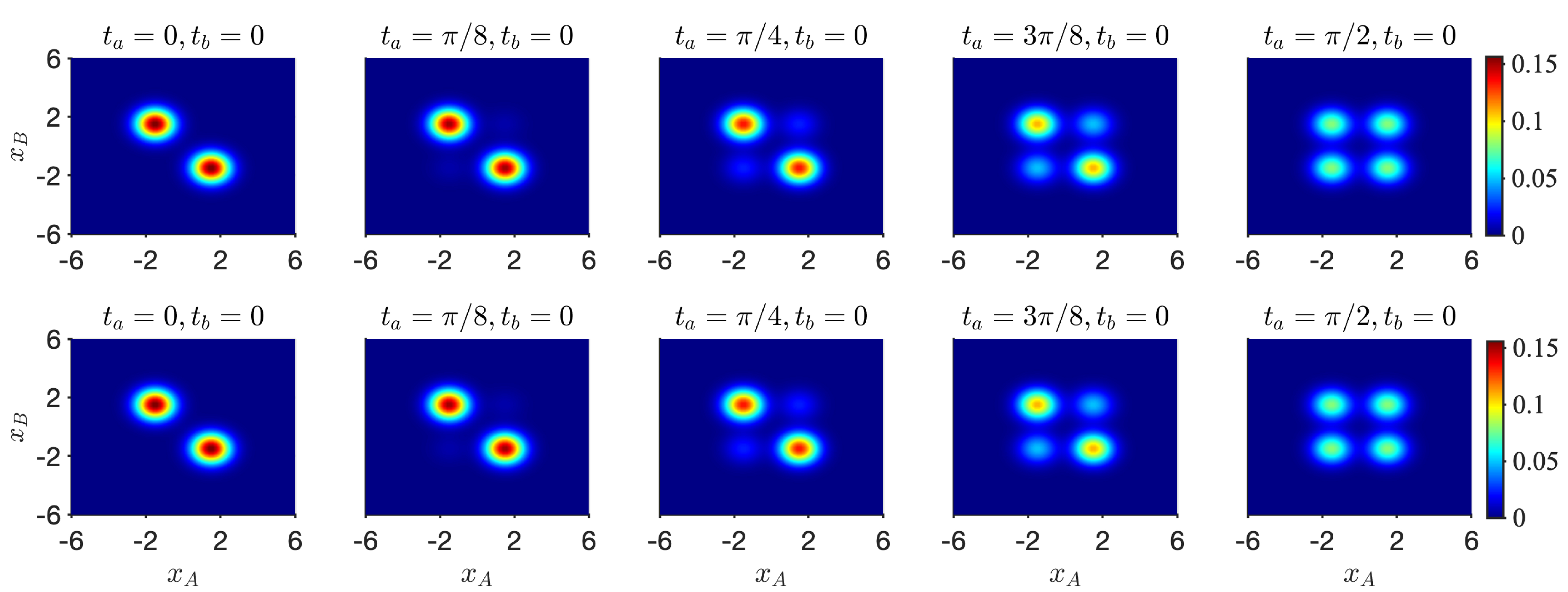}
\par\end{centering}
\caption{The timing of the final collapse stage of the measurement $S_{1}^{(B)}$
at time $t_{1}$ on $B$ is immaterial, for $\alpha,\beta>1$: 
The plots show a sequence of snapshots of $Q(x_{A},x_{B})$ as the
system described in Figure 14 evolves for the evaluation of $\langle S_{1}^{(A)}S_{2}^{(A)}\rangle$
and $\langle S_{1}^{(A)}S_{3}^{(A)}\rangle$.  (a) The top two sequences:
Here, $\alpha=\beta=0.5$. The top sequence shows the evolution if
no collapse of the measurement at $B$ has taken place before the
time $t_{3}=\pi/2\Omega$. The lower sequence shows the evolution
assuming the collapse has taken place at time $t_{b}=t_{1}=0$. (b)
The lower two sequences: As for (a), but here $\alpha=\beta=1.5$.
The states for the top and lower sequences although different are
visually indistinguishable.\textcolor{red}{}}
\end{figure*}
Now let us assume the final stage of the measurement at $B$ \emph{was}
made at the time $t=0$, meaning that the pointer measurement consisting
of a readout of the value of spin $S_{1}^{(B)}$ occurred at this
time. In this case, the system $B$ has been coupled to a third system,
so that a ``collapse'' occurs, for the system $B$ (and also for
$A$). The system immediately after $t=0$ is given as the mixture
$\rho_{mix}$ (eqn (\ref{eq:mixqstate})). The subsequent evolution
is different to the evolution of the superposition (\ref{eq:cat-1-3-1}).
If a measurement is made at the later time $t_{a}$, the joint probability
$P(X_{A},X_{B})_{mix}$ is readily calculated, as shown in the Appendix.
The two probabilities $P(X_{A},X_{B})_{sup}$ and $P(X_{A},X_{B})_{mix}$
are indeed different, due to the interference terms present for the
superposition state, where the collapse does not occur. However, for
$\alpha$ and $\beta$ large, the two probabilities are indistinguishable.

The result is illustrated in Figure 15 where we consider measurements
of $\langle S_{1}^{(A)}S_{2}^{(A)}\rangle$ and $\langle S_{1}^{(A)}S_{3}^{(A)}\rangle$,
for both small and large $\alpha$ and $\beta$. The Q functions
depicted in the final plots of the top and lower sequences directly
reflect the probabilities $P(X_{A},X_{B})_{sup}$ and $P(X_{A},X_{B})_{mix}$
for the two cases, where collapse has not, and has, taken place, respectively,
by the time of the detection at $A$. Although numerically different,
these plots are, in effect, indistinguishable, for $\alpha,\beta>1$.

A similar result occurs if we consider the measurement of $\langle S_{2}^{(A)}S_{3}^{(A)}\rangle$.
It makes no effective difference to the joint probabilities $P(X_{A},X_{B})$
whether or not the final collapse stage of the measurement $S_{2}^{(B)}$
at $B$ is delayed until a later time. The details are given in the
Appendix.\textcolor{blue}{}

\subsection{Interpreting the delayed-collapse cat-state experiment}

We ask what can be concluded from the Leggett-Garg-Bell experiments
described in Sections IV and V.A, where the collapse stage of the
measurement can be delayed? The latter will imply that one can perform
the macroscopic Bell and Leggett-Garg tests as delayed choice and
quantum eraser experiments \cite{delayed-choice-qubit}. This follows
naturally from the mapping from the microscopic to macroscopic qubits,
which is ideal in the limit of large $\alpha$, for the angle choices
needed for the delayed choice experiments. The new feature, compared
to former delayed-choice experiments, is that the results are at the
macroscopic level, where the qubit spin values reflect the macroscopically
distinct amplitudes.

An argument can be given for interpretating the delayed collapse gedanken
experiment in a way that is consistent with the premise of weak macroscopic
realism (wMR). Here, the definition of wMR includes the assumption
of macroscopic locality for the pointer (MLP) at A, as defined in
Section III.E. The argument is based on the macroscopic nature of
the experiment. After the time $t_{2}$, the final collapse stage
of the measurement $S_{2}^{(B)}$ at $B$ gives information about
the macroscopic value of the spin $S_{2}^{(A)}$ of system $A$ (should
it be measured). The delayed-collapse analysis of Sections V.A confirms
that this information can be revealed at system $B$ an infinite time
after $t_{2}$, and after further events at $A$. The system $B$
can be separated from $A$ by an infinite distance. A strong argument
can then be given that carrying out the collapse stage of the measurement
at $B$ does not \emph{macroscopically} change the state-description
of $A$ at the former time $t_{2}$ i.e. does not change the outcome
for the macroscopic spin $S_{2}^{(A)}$ at $A$.

We note that the wMR interpretation does \emph{not} mean that prior
to the measurement of spin $S_{2}^{(B)}$ at $B$ the system is in
the \emph{quantum} state $|\alpha\rangle$ or $|-\alpha\rangle$,
since these states are \emph{microscopically} specified, giving predictions
for all measurements that might be performed on $A$. That the system
in the cat state cannot be viewed as a classical mixture can be negated
in multiple ways, e.g. by observing fringes in the distribution for
the orthogonal quadrature $\hat{P}_{A}$ (for a single system $A$)
or by deducing an Einstein-Podolsky-Rosen (EPR) paradox for the entangled
state $|\psi_{Bell}\rangle$, along the lines given in Refs. \cite{eric_marg-1,irrealism-fringes,macro-coherence-paradox,macro-pointer,Bohm-1,epr-1,bohm-eric,epr-r2}.
We give an example of this in the next subsection.

The consistency with wMR is explained as follows. Suppose the systems
$A$ and $B$ are prepared at a time $t_{j}$ in a macroscopic superposition
$|\psi_{pointer}\rangle$ of states with definite outcomes for pointer
measurements $\hat{M}^{(A)}$ and $\hat{M}^{(B)}$, an example being
(as $\alpha,$$\beta\rightarrow\infty$)
\begin{equation}
|\psi_{Bell}\rangle=\mathcal{N}(|\alpha\rangle|-\beta\rangle-|-\alpha\rangle|\beta\rangle)\label{eq:cat-1-3-2}
\end{equation}
Here,  the states with definite outcomes for $\hat{M}^{(A)}$ and
$\hat{M}^{(B)}$ are the states with definite outcomes for the spin
$S_{j}^{(A)}$. The assumption of wMR asserts that the system $A$
at the time $t_{j}$ is in one or other of two macroscopically specified
states $\varphi_{+}$ and $\varphi_{-}$, for which the result of
a measurement of spin $S_{j}^{(A)}$ is deterministically predetermined
i.e. the system $A$ at time $t_{j}$ may be described by a macroscopic
hidden variable $\lambda_{j}^{(A)}$. For the bipartite system, wMR
is to be consistent with MLP. MLP asserts that the value of the macroscopic
hidden variable $\lambda_{j}^{(A)}$ for the system $A$ cannot be
changed by any spacelike separated event or measurement at the system
$B$ that takes place at time $t\geq t_{j}$ e.g. it cannot be changed
by a future event at $B$. The interpretation is that the system $A$
at each time $t_{1}$, $t_{2}$ and $t_{3}$ was in one or other of
states $\varphi_{j,+}$ or $\varphi_{j,-}$ with a definite value
of spin $S_{j}^{(A)}$, and that the failure of macrorealism arises
from the initial entanglement as the systems evolve dynamically at
the intermediate times.

The assumption of MLP is to be distinguished from the stronger assumption,
macroscopic locality (ML), introduced earlier, as part of the premise
of macroscopic local realism (MLR). The premise of ML assumes locality
to apply for spacelike separations where the measurement on system
$A$ can vary, so that $A$ is \emph{not necessarily prepared in the
pointer basis}. The above wMR interpretation is thus not contradicted
by the violation of macroscopic Bell inequalities. In fact, we have
seen in Section III that, consistent with the validity of wMR, there
is no violation of these Bell inequalities when there is no unitary
transformation $U_{A}$ at $A$. This is because in that case the
system is in (immediately prior to the measurements) a state $|\psi_{pointer}\rangle$
where wMR applies.

\subsection{Einstein-Podolsky-Rosen-type paradox based on macroscopic realism}

The EPR argument argues the incompleteness of quantum mechanics based
on the assumption of local realism, or local causality \cite{epr-1}.
One may also argue an EPR-type paradox for the cat states based on
the validity of weak macroscopic realism. Here, we apply the argument
given in Ref. \cite{macro-coherence-paradox}.

One considers the superposition state $c_{1}|\alpha\rangle+ic_{2}|-\alpha\rangle$
(for $\alpha$ large), similar to that prepared at the time $t_{2}$
in the Leggett-Garg-Bell tests. Here, $|c_{1}|^{2}+|c_{2}|^{2}=1$.
Macroscopic realism postulates that the system $A$ in such a state
is actually in one or other state $\varphi_{+}$ and $\varphi_{-}$
for which the value of the macroscopic spin $S_{2}^{(A)}$ is predetermined.
The spin $S_{2}^{(A)}$ is measured by the sign of $X_{A}$: the distribution
$P(X_{A})$ gives two separated Gaussian peaks, each with variance
$(\Delta X_{A})^{2}=1/2$.

One may specify the degree of predetermination of $X_{A}$ for the
macroscopic states, $\varphi_{+}$ and $\varphi_{-}$, by postulating
for each state $\varphi_{+}$ and $\varphi_{-}$ that the variance
be some specified value, which we take as $(\Delta X_{A})_{+}^{2}$
and $(\Delta X_{A})_{-}^{2}$ respectively. If we assume each of $\varphi_{+}$
and $\varphi_{-}$ to be \emph{quantum} states, then for each the
Heisenberg uncertainty relation implies $(\Delta X_{A})(\Delta P_{A})\geq1/2$.
We then argue, as was done in ref. \cite{macro-coherence-paradox},
that for the ensemble of systems in a classical mixture of such states,
$\varphi_{+}$ and $\varphi_{-}$, we require $(\Delta X_{A})_{ave}(\Delta P_{A})\geq1/2$,
where $(\Delta X_{A})_{ave}^{2}=P_{+}(\Delta X_{A})_{+}^{2}+P_{-}(\Delta X_{A})_{-}^{2}$
, $P_{+}=|c_{+}|^{2}$ and $P_{-}=|c_{-}|^{2}$. The observation of
\begin{equation}
\varepsilon_{M}\equiv(\Delta X_{A})_{ave}(\Delta P_{A})<1/2\label{eq:macro-epr}
\end{equation}
implies incompatibility of macroscopic realism with the completeness
of quantum mechanics, since the localised states $\varphi_{+}$ and
$\varphi_{-}$ cannot be given as quantum states.

If we take $(\Delta X_{A})_{+}^{2}=(\Delta X_{A})_{-}^{2}=1/2$, to
match that the states have the variances associated with the two Gaussian
peaks (the coherent states), we find the condition (\ref{eq:macro-epr})
is satisfied for $(\Delta P_{A})^{2}<1/2$. This is indeed the case
for the superposition above. The distribution $P(P_{A})$ is 
\begin{equation}
P(P_{A})=\frac{e^{-P_{A}^{2}}}{\sqrt{\pi}}\{1+\frac{1}{\sqrt{2}}\sin(2\sqrt{2}P_{A}|\alpha|)\}\label{eq:supfringep-1}
\end{equation}
which shows fringes, as given in ref. \cite{yurke-stoler-1}. The
variance is \textcolor{red}{}$(\Delta P_{A})^{2}=\frac{1}{2}-\alpha^{2}e^{-4\alpha^{2}}$
\cite{macro-pointer}, indicating a paradox.

\section{Conclusion}

In this paper, we have provided ways to test quantum mechanics against
macroscopic realism using cat states. We consider two definitions
of macroscopic realism. The first, macrorealism (M-R), supplements
the assumption of macroscopic realism with the premise of macroscopic
noninvasive measurability. The second, macroscopic local realism (MLR),
supplements with the premise of macroscopic locality. In this paper,
``macroscopically distinct'' refers to separations well beyond the
quantum noise level $\hbar$.

In Sections II-IV, we show that quantum mechanics predicts failure
of MLR (and M-R) for two spacelike-separated systems prepared in entangled
cat states, in a way that is directly analogous to the original Bell
and Leggett-Garg proposals. To test MLR, we consider CHSH-type
Bell inequalities and three-time Leggett-Garg-Bell inequalities, where
measurements are made at two spacelike separated locations, at successive
times. One may construct similar tests, based on the Leggett-Garg-Bell
inequality described in Section II, involving four times.

The tests give a convincing strategy to demonstrate the non-invasiveness
of a measurement that is assumed in a Leggett-Garg test of macro-realism.
For a macroscopic system, the usual argument is that the disturbance
(at least for some ideal measurement) becomes vanishingly small, so
as to have a negligible affect on the later dynamics. In order to
refute the criticism that there \emph{is} a disturbance, methods have
been developed to quantify the effect of such a disturbance, if macroscopic
realism is satisfied. One approach performs a control experiment
that prepares two macroscopically distinguishable states ($\psi_{1}$
and $\psi_{2}$) and then quantifies the effect of disturbance on
the inequality, if the system is indeed in one of those states \cite{NSTmunro-1}.
A counterargument however is that the system at time $t_{2}$ is
in a state microscopically different to $\psi_{1}$ (or $\psi_{2}$),
a state which was not prepared in the laboratory. In the present
paper, such criticisms are avoided, because macroscopic realism is
defined more broadly. One does not assume that the macroscopically
distinct states are specific\emph{ }states $\psi_{1}$ and $\psi_{2}$,
nor that the states are quantum states. A direct disturbance due to
measurement is ruled out, because the result for the measurement of
$A$ is inferred from the measurement on the separated system $B$.
The non-classicality is then explained by quantum nonlocality,
but at a macroscopic level.

In Section V, we have extended the bipartite Leggett-Garg analysis,
to demonstrate that a delay of the final irreversible stage of the
measurement makes no difference to the predictions, provided the amplitude
of the coherent states is large. This implies that one could perform
delayed choice experiments, similar to those examined for qubits in
Refs. \cite{delayed-choice-qubit,delayed-choice-interpretation}.
The possibility of delayed choice experiments is expected because
of the mapping onto the qubit system, for large $\alpha$, $\beta$
(as the coherent states become orthogonal), for the rotation angles
required for the qubit experiments.  The new feature is the
observation at a macroscopic level, since the qubits correspond to
macroscopically distinct coherent states. As with the qubit experiments,
this does not however imply acausal effects \cite{delayed-choice-interpretation}.

It is clear from the violations of the macroscopic inequalities presented
in this paper that MLR, macroscopic local causality (MLC), and M-R
fail. We have explained in Section II that the MLR is defined as deterministic
macroscopic (local) realism (dMR), that the system is predetermined
to be in a state giving a definite outcome for the pointer measurement
($\hat{M}_{\theta}$ or $\hat{M}_{\theta'}$) prior to the unitary
rotation $U$ that determines the measurement setting, $\theta$ or
$\theta'$. The violation of the macroscopic Bell inequalities given
in this paper show that dMR fails.

The results are however consistent with an interpretation in which
a \emph{weak macroscopic realism} (wMR) holds. The initial Bell state
gives predictions for joint probabilities $P(X_{A},P_{A},X_{B},P_{B})$
at time $t_{1}$ that are only microscopically different (of order
$\hbar e^{-|\alpha|^{2}}$) to those of a non-entangled mixture. Yet,
after the choice of measurement setting, the final joint probabilities
are \emph{macroscopically} different. In the wMR interpretation, the
macroscopic differences arise during the unitary dynamics, corresponding
in a Bell experiment to the choice of measurement setting. Such dynamics
shifts the system into a superposition $|\psi_{pointer}\rangle$ of
states with definite outcomes for the measurement of the macroscopic
spin $S_{i}^{(A)}$, at a given time $t_{i}$. The assumption of wMR
postulates that a system in a superposition $|\psi_{pointer}\rangle$
is describable by a macroscopic ``element of reality'', $\lambda_{M}$,
which predetermines the result of the pointer measurement for the
macroscopic qubit value $S_{i}^{(A)}$. The interpretation is that
the system is in one or other states $\varphi_{+}$ and $\varphi_{-}$
giving a definite value for the macroscopic pointer measurement.

An EPR-type paradox arises from the assumption of weak macroscopic
realism, if correct. The element of reality ``state'', $\varphi_{+}$
or $\varphi_{-}$, of the system prior to the measurement cannot be
a \emph{quantum} state. This is evident from calculations given in
\cite{macro-coherence-paradox} and explained in section V.C, where
the states $\varphi_{+}$ or $\varphi_{-}$ that would apply to the
cat state $|\alpha\rangle+|-\alpha\rangle$ can be shown to be inconsistent
with the uncertainty principle. Similar paradoxes have been illustrated
for the two-slit experiment using the concept of irrealism \cite{irrealism-fringes},
and for entangled cat states \cite{eric_marg-1}, based on the logic
of the original EPR paradoxes that reveal the inconsistency of the
completeness of quantum mechanics with local realism \cite{epr-1,Bohm-1,bohm-eric,epr-r2}.

Finally, we comment on the possibility of an experiment. Entangled
cat states have been generated \cite{cat-bell-wang-1}. The challenge
is to realise the unitary rotation, given by the nonlinear Hamiltonian
which has a quartic dependence on the field intensity. This is necessary
for the full Bell experiment. However, a macroscopic delayed choice/
quantum eraser experiment may be carried out more straightforwardly,
since this depends only on the creation of a simple cat state superposition,
which can be realised at times $t=\pi/2\Omega$ from the nonlinear
Kerr interaction $H_{NL}=\Omega\hat{n}^{2}$ \cite{yurke-stoler-1,manushan-cat-lg}.
This interaction has been experimentally achieved \cite{collapse-revival-bec-2,collapse-revival-super-circuit-1}.

\section*{Acknowledgements}

This research has been supported by the Australian Research Council
Discovery Project Grants schemes under Grant DP180102470. 

\begin{widetext}

\section*{Appendix}

\subsection{Calculation of Bell violations}

The joint probability is $P(X_{A},X_{B})=|\langle X_{A}|\langle X_{B}|\psi_{f}\rangle|^{2}$
where $|X_{A}\rangle$, $|X_{B}\rangle$ are eigenstates of $X_{A}$
and $X_{B}$ respectively. Using the  overlap  $\langle x|e^{i\theta}\alpha\rangle=\frac{1}{\pi^{1/4}}\exp(-\frac{x^{2}}{2}+\frac{2x|\alpha|e^{i\theta}}{\sqrt{2}}-\{\frac{|\alpha|e^{i\theta}}{\sqrt{2}}\}^{2}-\frac{|\alpha|^{2}}{2})$,
we find
\begin{align}
\langle X_{A}|\langle X_{B}|\psi_{f}\rangle & =\frac{\exp^{-(|\alpha|^{2}+|\beta|^{2})}}{\sqrt{2\pi}}\{\exp^{-\frac{X_{A}^{2}}{2}+\sqrt{2}X_{A}|\alpha|}[b\exp^{-\frac{X_{B}^{2}}{2}-\sqrt{2}X_{B}|\beta|}-i\bar{b}\exp^{-\frac{X_{B}^{2}}{2}+\sqrt{2}X_{B}|\beta|}]\nonumber \\
 & -\exp^{-\frac{X_{A}^{2}}{2}-\sqrt{2}X_{A}|\alpha|}[b\exp^{-\frac{X_{B}^{2}}{2}+\sqrt{2}X_{B}|\beta|}+i\bar{b}\exp^{-\frac{X_{B}^{2}}{2}-\sqrt{2}X_{B}|\beta|}]\}\label{eq:pf-1}
\end{align}

\subsection{Interference effects}

Where $t_{a}=0$, the joint probability $P(X_{A},X_{B})$ for outcomes
$X_{A}$ and $X_{B}$ for the system prepared in the entangled superposition
(\ref{eq:cat-1-3}) is 
\begin{eqnarray}
P(X_{A},X_{B}) & = & \mathcal{N}^{2}\{|\langle X_{A}|\alpha\rangle\langle x_{B}|U^{(B)}(t_{b})|-\beta\rangle|^{2}+|\langle X_{A}|-\alpha\rangle\langle X_{B}|U^{(B)}(t_{b})|\beta\rangle|^{2}\nonumber \\
 &  & +\langle X_{A}|\alpha\rangle\langle X_{B}|U^{(B)}(t_{b})|-\beta\rangle\langle X_{A}|-\alpha\rangle^{*}\langle X_{B}|U^{(B)}(t_{b})|\beta\rangle^{*}\nonumber \\
 &  & +\langle X_{A}|\alpha\rangle^{*}\langle X_{B}|U^{(B)}(t_{b})|-\beta\rangle^{*}\langle X_{A}|-\alpha\rangle\langle X_{B}|U^{(B)}(t_{b})|\beta\rangle\}\label{eq:pxxint-1}
\end{eqnarray}
This contains an interference term proportional to $\exp(-X_{A}^{2}-2|\alpha|^{2})$
that vanishes with large $\alpha$. The expression reduces to
the result for the mixture $\rho_{mix}$, given by
\begin{eqnarray}
P(X_{A},X_{B}) & = & \frac{1}{2}\{|\langle X_{A}|\alpha\rangle\langle X_{B}|U_{2}(t_{b})|-\beta\rangle|^{2}+|\langle X_{A}|-\alpha\rangle\langle X_{B}|U_{2}(t_{b})|\beta\rangle|^{2}\}\label{eq:pxxmix-1}\\
\nonumber 
\end{eqnarray}
because the interference term contains the expression $\langle X_{A}|\alpha\rangle\langle X_{A}|-\alpha\rangle^{*}=\frac{1}{\pi^{1/2}}\exp(-X_{A}^{2}-2|\alpha|^{2})$
which is small for orthogonal state, with large $\alpha_{0}$.

By contrast, for $t_{a}\neq0$ and $t_{b}\neq0$, we compare the
expressions $P(X_{A},X_{B})$ for the evolved cat state given by $|\psi_{f}\rangle$
of eq. (\ref{eq:234-1-1-1}), with that of the evolved mixture 
\begin{eqnarray}
\rho_{mix}(t_{a},t_{b}) & = & \frac{1}{2}\Bigl(U^{(B)}(t_{b})|-\beta\rangle U^{(A)}|\alpha\rangle\langle\alpha|U^{(A)\dagger}\langle-\beta|U^{(B)\dagger}(t_{b})\nonumber \\
 &  & +U^{(B)}|\beta\rangle U^{(A)}|-\alpha\rangle\langle-\alpha|U^{(A)\dagger}\langle\beta|U^{(B)\dagger}\Bigl)\label{eq:rho-mix-evolve}
\end{eqnarray}
The expressions are the same apart from interference terms, such
as
\begin{equation}
\langle X_{A}|U^{(A)}(t_{a})|\alpha\rangle\langle X_{B}|U^{(B)}(t_{b})|-\beta\rangle\times\langle X_{A}|U^{(A)}(t_{a})|-\alpha\rangle^{*}\langle X_{B}|U^{(B)}(t_{b})|\beta\rangle^{*}\label{eq:54}
\end{equation}
We see from the solutions given by eqs. (\ref{eq:state3}-\ref{eq:state5})
that for the suitable choices of $t_{a}$ , $U^{(A)}(t_{a})|-\alpha\rangle$
(for example) may involve terms such as $|\alpha\rangle$, and $U^{(A)}(t_{a})|\alpha\rangle$
involves terms in $|\alpha\rangle$. This then leads to a contribution
of type $\langle X_{A}|\alpha\rangle\langle X_{A}|\alpha\rangle^{*}=\frac{1}{\pi^{1/2}}\exp(-X_{A}^{2}+2\sqrt{2}X_{A}\alpha-2|\alpha|^{2})$\textcolor{red}{{}
}and therefore significant interference terms (when $X_{A}\sim\sqrt{2}\alpha$).
These terms are the origin of the macroscopic difference between the
probability distributions in Figures 6 for nonzero $t_{a}$ and $t_{b}$,
which leads to the macroscopic Bell violation.

\subsection{Leggett-Garg test for a single system: Q function dynamics}

The Q function at time $t_{1}$ is that of a coherent state $|\alpha_{0}\rangle$
where we take $\alpha_{0}$ to be real.\textcolor{red}{}
\begin{equation}
Q(x_{A},p_{A},t_{1})=\frac{e^{-p_{A}^{2}}}{\pi}e^{-(x_{A}-\alpha_{0})^{2}}\label{eq:qcoh}
\end{equation}
At time $t_{2}$, the system is in the superposition state (\ref{eq:supt2})
which has the Q function
\begin{align}
Q(x_{A},p_{A},t_{2})= & \frac{e^{-p_{A}^{2}}}{2\sqrt{2}\pi}\{c_{+}e^{-(x_{A}-\alpha_{0})^{2}}+c_{-}e^{-(x_{A}+\alpha_{0})^{2}}-2e^{-x_{A}^{2}}e^{-\alpha_{0}^{2}}\sin(2p_{A}\alpha_{0})\}\label{eq:q2}
\end{align}
where $c_{\pm}=\sqrt{2}\pm1$. \textcolor{red}{}\textcolor{blue}{\emph{}}\textcolor{red}{}Assuming
no measurement takes place at time $t_{2}$, the system at time $t_{3}$
is in the superposition state (\ref{eq:supt3}) with\textcolor{red}{}
\begin{align}
Q(x_{A},p_{A},t_{3})= & \frac{e^{-p_{A}^{2}}}{2\pi}\{e^{-(x_{A}-\alpha_{0})^{2}}+e^{-(x_{A}+\alpha_{0})^{2}}-2e^{-x_{A}^{2}}e^{-\alpha_{0}^{2}}\sin(2p_{A}\alpha_{0})\}\label{eq:q3}
\end{align}
If the measurement is performed at $t_{2}$, the system ``collapses''
to either $|\alpha_{0}\rangle$ (with probability $\cos^{2}\pi/8$)
or $|-\alpha_{0}\rangle$ (with probability $\sin^{2}\pi/8$) and
then evolves at time $t_{3}$ respectively to either $U_{\pi/8}|\alpha\rangle$
or $U_{\pi/8}|-\alpha\rangle$ as given by eqn. (\ref{eq:supt2}).

There are two cases to compare: whether or not a measurement is performed
at time $t_{2}$. Figure 12  shows the sequence for the Q functions
at the times $t_{1}$ , $t_{2}$ and $t_{3}$ for these two cases.
The top sequence, modelling where a measurement is not performed,
shows sequentially the three Q functions, (\ref{eq:qcoh}), (\ref{eq:q2})
and (\ref{eq:q3}). The lower sequence plots where a measurement
is performed at time $t_{2}$, assuming this is done in such a way
to instigate a collapse into one or other of the coherent states,
as described above. The second plot of the lower sequence is therefore
 the Q function\textcolor{red}{}
\begin{align}
Q(x_{A},p_{A},t_{2}) & =\frac{e^{-p_{A}^{2}}}{2\sqrt{2}\pi}\{c_{+}e^{-(x_{A}-\alpha_{0})^{2}}+c_{-}e^{-(x_{A}+\alpha_{0})^{2}}\}\nonumber \\
\label{eq:q2-2}
\end{align}
representing the average state $\rho_{mix}(t_{2})$ of the system
at time $t_{2}$, immediately after the measurement at time $t_{2}$:
$\rho_{mix}(t_{2})=\cos^{2}\pi/8|\alpha_{0}\rangle\langle\alpha_{0}|+\sin^{2}\pi/8|-\alpha_{0}\rangle\langle-\alpha_{0}|$.
 The Q function for the final state at time $t_{3}$, if the measurement
has taken place at time $t_{2}$, is given by the evolution of $\rho_{mix}(t_{2})$
for the time $\pi/4\Omega$. The final \emph{average} state is $\rho_{mix}(t_{3})=\cos^{2}\pi/8|\psi_{+}\rangle\langle\psi_{+}|+\sin^{2}\pi/8|\psi_{-}\rangle\langle\psi_{-}|$
where $|\psi_{\pm}\rangle=U_{\pi/8}|\pm\alpha\rangle$. The Q function
is\textcolor{red}{\emph{}}
\begin{align}
Q(x_{A},p_{A},t_{3})= & \frac{e^{-p_{A}^{2}}}{8\pi}\{6e^{-(x_{A}-\alpha_{0})^{2}}+e^{-(x_{A}+\alpha_{0})^{2}}-4e^{-x_{A}^{2}}e^{-\alpha_{0}^{2}}\sin(2p_{A}\alpha_{0})\}\label{eq:q3mix}
\end{align}
which is plotted as the third function of the lower sequence.\textcolor{red}{}

\subsection{Bipartite Leggett-Garg tests: correlation between the outcomes at
the different sites}

Here, we evaluate $B_{lg}=-\{\langle S_{1}^{(B)}S_{2}^{(A)}\rangle-\langle S_{1}^{(B)}S_{3}^{(A)}\rangle+\langle S_{2}^{(B)}S_{3}^{(A)}\rangle\}$
and the value of the conditional probabilities $P(S_{i}^{(A)}=1|S_{i}^{(B)}=-1)$
where $i=1,2$ for the case $\alpha=\beta$. These quantities are
worked out exactly as one would measure them. We first evaluate $P(X_{A},X_{B})$
at time $t_{2}$.\textcolor{red}{{} }
\begin{align}
P(X_{A},X_{B}) & =\left|\langle X_{B}|\langle X_{A}|U_{\pi/8}^{(A)}U_{\pi/8}^{(B)}|\psi_{Bell}\rangle\right|^{2}\nonumber \\
 & =2\frac{e^{-X_{A}^{2}-X_{B}^{2}-2\left|\alpha\right|^{2}-2\left|\beta\right|^{2}}}{\pi{\color{blue}{\normalcolor \left(1-e^{-2\left|\alpha\right|^{2}-2\left|\beta\right|^{2}}\right)}}}\sinh^{2}(\sqrt{2}X_{A}\left|\alpha\right|-\sqrt{2}X_{B}\left|\beta\right|)\label{eq:ap1}
\end{align}
where we use
\begin{align}
\langle X_{B}|\langle X_{A}|U_{\pi/8}^{(A)}U_{\pi/8}^{(B)}|\psi_{Bell}\rangle & =\mathcal{N}e^{-i\pi/4}\left(\langle X_{A}|\alpha\rangle\langle X_{B}|-\beta\rangle-\langle X_{A}|-\alpha\rangle\langle X_{B}|\beta\rangle\right)\nonumber \\
 & =2\mathcal{N}\frac{e^{-\frac{X_{A}^{2}}{2}-\frac{X_{B}^{2}}{2}}}{\pi^{1/2}}e^{-\left|\alpha\right|^{2}-\left|\beta\right|^{2}-i\pi/4}\sinh(\sqrt{2}X_{A}\left|\alpha\right|-\sqrt{2}X_{B}\left|\beta\right|)\label{eq:ap2}
\end{align}
In fact, because the system at time $t_{2}$ remains in a Bell state
$|\psi_{Bell}\rangle$ (apart from a phase factor), this also represents
$P(X_{A},X_{B})$ for the time $t_{1}$. We note that
\begin{align}
P(X_{B}) & =\int P(X_{A},X_{B})dX_{A}=\frac{e^{-X_{B}^{2}-2\left|\alpha\right|^{2}-2\left|\beta\right|^{2}}}{\sqrt{\pi}{\color{blue}{\normalcolor \{1-e^{-2\left|\alpha\right|^{2}-2\left|\beta\right|^{2}}\}}}}\left(e^{2|\alpha|^{2}}\cosh(2\sqrt{2}|\beta|X_{B})-1\right)\label{eq:ap2-1}
\end{align}
which on integration gives $P(X_{B}>0)=1/2$. For the value of the
conditional probability $P(S_{i}^{(A)}=1|S_{i}^{(B)}=-1)$, $i=1,2$,
one measures

\textcolor{black}{
\begin{align}
{\color{black}{\color{blue}}P(X_{A}>0|X_{B}\leq0)} & {\color{black}=\frac{\int_{-\infty}^{0}\int_{0}^{\infty}P(X_{A},X_{B})dX_{A}dX_{B}}{\int_{-\infty}^{0}P(X_{B})dX_{B}}{\color{black}={\color{black}\frac{1}{2}+\frac{erf(\sqrt{2}\alpha)\times erf(\sqrt{2}\beta)}{{\color{black}2\left(1-e^{-2\alpha^{2}-2\beta^{2}}\right)}}}}}\label{eq:ap3}
\end{align}
}Similarly, we evaluate\textcolor{red}{}\textcolor{black}{
\begin{align}
{\color{black}}{\normalcolor }{\normalcolor }{\normalcolor }\langle S_{1}^{(B)}S_{2}^{(A)}\rangle & =-\frac{\sqrt{2}erf(\sqrt{2}\alpha)erf(\sqrt{2}\beta)}{2{\color{black}{\color{blue}{\color{black}\{1-e^{-2\left|\alpha\right|^{2}-2\left|\beta\right|^{2}}\}}}}}\nonumber \\
\langle S_{1}^{(B)}S_{3}^{(A)}\rangle & =0\label{eq:ap4-3}
\end{align}
where $\langle S_{2}^{(B)}S_{3}^{(A)}\rangle=\langle S_{1}^{(B)}S_{2}^{(A)}\rangle$.
This gives  the result (\ref{eq:B-lg}). The plots of Figure 13
and 15 reveal that for $\alpha=\beta>1$, the positive and negati}ve
values for the outcomes of $X_{A}$ correspond to distinct macroscopically
separated Gaussian peaks in the different quadrants. The conditional
probability goes to $1$ as $\alpha$ is larger, which justifies that
the measurement of $S_{i}^{(B)}$at $B$ indicates the value $S_{i}^{(A)}$at
$A$.

\subsection{Delayed Collapse for measurement $\langle S_{1}^{(A)}S_{2}^{(A)}\rangle$}

The state formed after the evolution at $A$ is the superposition
\begin{equation}
U_{A}(t_{a})|\psi_{Bell}\rangle=\mathcal{N}\{U_{A}(t_{a})|\alpha\rangle_{a}|-\beta\rangle_{b}-U_{A}(t_{a})|-\alpha\rangle_{a}|\beta\rangle_{b}\}\label{eq:cat-1-3-1-2}
\end{equation}
The final joint distribution at time $t_{a}$ is (assuming there has
been no prior collapse at $B$)
\begin{eqnarray}
P(X_{A},X_{B})_{sup} & = & \mathcal{N}^{2}(\langle X_{A}|U_{A}(t_{a})|\alpha\rangle_{a}\langle X_{B}|-\beta\rangle_{b}-\langle X_{A}|U_{A}(t_{a})|-\alpha\rangle_{a}\langle X_{B}|\beta\rangle_{b})\nonumber \\
 &  & \times(\langle\alpha|_{a}U_{A}^{\dagger}(t_{a})|X_{A}\rangle\langle-\beta|_{b}X_{b}\rangle-\langle-\alpha|_{a}U_{A}^{\dagger}(t_{a})|X_{A}\rangle\langle\beta|_{b}X_{b}\rangle)\label{eq:prob1-1}
\end{eqnarray}
Now let us assume the final stage of the measurement at $B$ \emph{was}
made at the time $t=0$, meaning that the pointer measurement consisting
of a readout of the value of spin $S_{1}^{(B)}$ occurred at this
time. In this case, the system $B$ has been coupled to a third system,
so that a ``collapse'' occurs, for the system $B$ (and also for
$A$). The system (or rather, an ensemble of them) immediately after
$t=0$ is given as the mixture $\rho_{mix}$ (eqn (\ref{eq:mixqstate})).
The system then evolves from a state that is \emph{either} $|\alpha\rangle|-\beta\rangle$
or $|-\alpha\rangle|\beta\rangle$, which is different to the evolution
of the superposition (\ref{eq:cat-1-3-1-2}). If a measurement is
made at the later time $t_{a}$, the joint probability is 
\begin{eqnarray}
P(X_{A},X_{B})_{mix} & = & \frac{1}{2}(\langle X_{A}|U_{A}(t_{a})|\alpha\rangle_{a}\langle X_{B}|-\beta\rangle_{b}\langle\alpha|_{a}U_{A}^{\dagger}(t_{a})|X_{A}\rangle\langle-\beta|_{b}X_{b}\rangle\nonumber \\
 &  & +\langle X_{A}|U_{A}(t_{a})|-\alpha\rangle_{a}\langle X_{B}|\beta\rangle_{b})\langle-\alpha|_{a}U_{A}^{\dagger}(t_{a})|X_{A}\rangle\langle\beta|_{b}X_{b}\rangle)\label{eq:prob2-1}
\end{eqnarray}
The two probabilities $P(X_{A},X_{B})$ are indeed different. However,
for $\alpha$ and $\beta$ large, examination shows that the two probabilities
(\ref{eq:prob1-1}) and (\ref{eq:prob2-1}) become indistinguishable\emph{.}

\subsection{Delayed collapse for measurement $\langle S_{2}^{(A)}S_{3}^{(A)}\rangle$}

We may also examine where a measurement is to be made at time $t_{2}$,
as in the evaluation of $\langle S_{2}^{(A)}S_{3}^{(A)}\rangle$.
Again, it makes no difference as to the evaluation, if the ``collapse''
stage is delayed until a later time. Wishing to measure the spin
of subsystem $B$ at time $t_{2}$ means we then ``stop'' the evolution
of system $B$ at $t_{b}=t_{2}$. If the system $A$ continues to
evolve, then the state of the system is ($t'_{a}=t_{A}-t_{2}$)
\begin{eqnarray}
U^{(A)}(t'_{A})|\psi_{Bell}\rangle & = & N_{\pm}e^{-i\pi/4}(U^{(A)}(t'_{A})|\alpha\rangle|-\beta\rangle-U^{(A)}(t'_{A})|-\alpha\rangle|\beta\rangle)\label{eq:bellunitary-2-1}
\end{eqnarray}
The final distributions $P(X_{A},X_{B})_{sup}$ and $P(X_{A},X_{B})_{mix}$
are similar to those above, replacing $t_{A}$ by $t'_{A}$. The lower
sequence of Figure 15 shows the Q function dynamics for the evolution
where there has been no collapse at time $t_{2}$. When plotted, the
sequence where a collapse has occured shows, even for moderate $\beta$,
no distinguishable difference, in either the final measured probabilities
or the Q function.

\end{widetext}

\end{document}